\documentstyle[psfig]{mn}

\newif\ifAMStwofonts

\newcommand{\lapp}{\mbox{\raisebox{-0.3em}{$\stackrel{\textstyle <}{\sim}$}}}
\newcommand{\gapp}{\mbox{\raisebox{-0.3em}{$\stackrel{\textstyle >}{\sim}$}}}

\title{NGC~4438 and its environment at radio wavelengths}
\author[Ananda Hota, D.J. Saikia and Judith A. Irwin]
    {Ananda Hota$^{1,2,3}$\thanks{hota@asiaa.sinica.edu.tw (AH), djs@ncra.tifr.res.in (DJS), 
     irwin@astro.queensu.ca (JAI)}, 
     D.J. Saikia$^2$ and Judith A. Irwin$^4$ \\  
$^1$ Joint Astronomy Programme, Indian Institute of Science, Bangalore 560 012, India \\
$^2$ National Centre for Radio Astrophysics, TIFR, Pune University Campus, Post Bag 3, Pune 411 007, India \\
$^3$ Institute of Astronomy and Astrophysics, Academia Sinica, P.O. Box 23-141, Taipei 106, Taiwan \\
$^4$ Department of Physics, Queen's University, Kingston K7L 3N6, Canada }

\date{Accepted.    Received Dec. }

\pagerange{\pageref{firstpage}--\pageref{lastpage}}
\pubyear{  }

\begin{document}

\maketitle

\begin{abstract}
We present multi-frequency radio-continuum and H{\sc i} observations of NGC~4438,
the highly-disturbed, active galaxy in the Virgo cluster, with the Very Large Array
(VLA) and the Giant Metrewave Radio Telescope (GMRT).
High-resolution observations of the central 1~kpc with the VLA at 4860 and
8460 MHz show the presence of an inverted-spectrum radio nucleus
located between the highly asymmetric lobes of radio emission. This demonstrates
that these lobes which are seen in radio continuum, H$\alpha$ and x-ray wavelengths
and are located at $\sim$230 and 730 pc from the nucleus arise due to an active
galactic nucleus (AGN) rather than a compact nuclear starburst. The low-frequency radio
continuum observations made with the GMRT detect
the extended emission on the western side of the galaxy whose spectral index
is flatter at higher frequencies and suggests that it is a mixture of thermal and
non-thermal emission. 

The H{\sc i} observations show an elongated structure which is displaced
by $\sim$4.1kpc on the western side of NGC~4438 and has a size of
$\sim$9.8 kpc and a mass of 1.8$\times$10$^8$M$_\odot$. The velocity field 
suggests systematic rotation, consistent with earlier observations. 
These observations also detect
H{\sc i} emission from the disk of the galaxy with a mass of 1.2$\times$10$^8$M$_\odot$.
We detect a faint H{\sc i}-tail towards the north of NGC~4438 close to a stellar
tail seen earlier in deep optical observations. This H{\sc i}-tail has a total extent of
$\sim$50 kpc and a mass of 1.4$\times$10$^8$ M$_\odot$ if it is at the distance
of NGC~4438.  The  velocity of the H{\sc i} tail is $\sim$$-$10 km s$^{-1}$
similar to that of H{\sc i} emission from IC~3355, but the possibility that the tail could
be foreground Galactic emission cannot be ruled out. We discuss the different structures 
in the light of different models which have been suggested for this disturbed galaxy,
namely ram pressure stripping, tidal and ISM-ISM interactions.
\end{abstract}


\begin{table*}
 \centering
 \begin{minipage}{140mm}
  \caption{Basic data on NGC~4438.$^a$}
  \begin{tabular}{@{}cccccccccc@{}}
\hline
 RA (J2000)$^b$
& DEC (J2000)$^b$ & Type$^c$ &
 a $\times$ b$^d$ & PA$^e$ & i$^f$ & V$_{sys}$$^g$ & D$^h$ & q-factor$^i$ &  \hbox{H\,{\sc i}}$^j$ \\
 (h m s) & ($^\circ$ $^{\prime}$ $^{\prime\prime}$) &
& ($^\prime$ $\times$ $^\prime$) & $^\circ$ & $^\circ$ & (km s$^{-1}$) & (Mpc) & & deficiency\\
\hline
 12 27 45.67 & +13 00 31.5 & SA(s)0/a pec &
8.5 $\times$ 3.2 & 29 & 80 & 71$\pm$3 & 17 & 1.83 & $>$1.0\\
\hline
\end{tabular}\hfill\break
$a$ Taken from the NASA Extragalactic Database (NED) unless stated otherwise.  \hfill\break
$b$ Position of the radio nucleus from our high-resolution, VLA A-array image at 8460 MHz. \hfill\break
$c$ Morphological type.\hfill\break
$d$ Optical major and minor axes.\hfill\break
$e$ Position angle (PA) of major axis from Kenney et al. (1995). \hfill\break
$f$ Inclination angle from Kenney et al. (1995). \hfill\break
$g$ Heliocentric systemic velocity. \hfill\break
$h$ Distance of NGC~4438 taken drom Vollmer et al. (2005). For this distance 1$^{\prime\prime}$=82 pc. \hfill\break
$i$ q-factor: logarithmic ratio of FIR to radio luminosity (Reddy \& Yun 2004). \hfill\break
$j$ H\,{\sc i} deficiency parameter from Cayatte et al. (1994).
\end{minipage}
\end{table*}

\begin{figure}
\hbox{
  \psfig{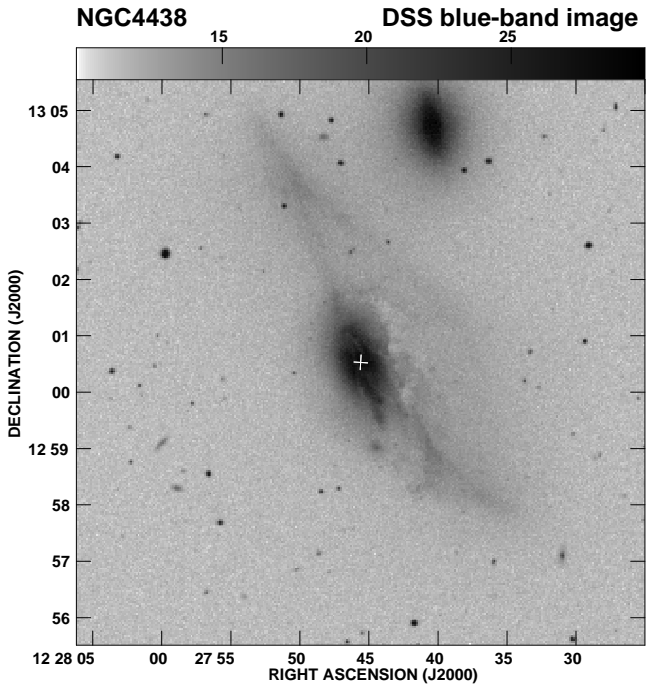}
   }
\caption[]{The DSS blue-band image of NGC~4438 and its companion, NGC~4435. The $+$
           sign marks the position of the radio nucleus discussed in this paper.}
\end{figure}

\section{Introduction}
The galaxy, NGC~4438 (VV188, Arp 120), which is located
only about 1$^\circ$ ($\sim$300 kpc)  from the centre of the  
Virgo cluster, is highly inclined (80$^\circ$) and has 
a very disturbed disk (Fig.~1)
with various components of the interstellar 
medium (ISM) being visible on the western side of the disk.
The basic properties of the galaxy are summarised in Table~1.
High-resolution x-ray observations with the Chandra telescope show 
emission from a $\sim$700 pc nuclear region, a $\sim$2.3 kpc spherical 
bulge and a network of filaments extending 4$-$10 kpc to the west 
and south-west of the galaxy (Machacek, Jones \& Forman 2004). These regions 
are well correlated with similar features seen in H$\alpha$
(Kenney \& Yale 2002; Chemin et al. 2005). Diffuse extended radio emission at 1.4 GHz
has been seen extending up to  $\sim$10 kpc on the western side of the galaxy. 
Some have suggested that these features on the western side might be due to interactions with the
intracluster medium (ICM) of the Virgo cluster since the relative velocity
of NGC~4438 is $\sim$1000 km s$^{-1}$.  A tidal encounter with the SB0 galaxy
NGC~4435 at a projected distance of $\sim$4.3 arcmin ($\sim$20 kpc)
is also likely to have affected NGC~4438. Kenney et al. (1995)
have suggested that most of the features of the disturbed ISM
are likely due to a high-velocity ISM-ISM collision between
NGC~4438 and 4435. 
Numerical simulations suggest that the centres of these
two galaxies have passed within 5$-$10 kpc of each other $\sim$10$^8$ yr ago
(Vollmer et al. 2005). Radio continuum observations of the nuclear region 
by Hummel \& Saikia (1991) have shown two well-defined lobes of radio emission 
in this galaxy which is unusual, though not unique, for a spiral galaxy. These lobes
are highly asymmetric with the north-western one being much brighter than the 
south-eastern one. x-ray and H$\alpha$+[N{\sc ii}] emission are also seen from 
these two lobes which exhibit a bubble-like structure and similar asymmetry in location,
brightness and size at these wavelengths (Kenney \& Yale 2002; Machacek et al. 2004). 

In this paper, we first present our results of radio continuum observations of the nuclear 
region and extended emission on the western side of the galaxy (Section 3). 
We then present the results of \hbox{H\,{\sc i}} observations with the GMRT and 
the VLA D-array (Section 4).
This is followed by a brief discussion and a summary of the results (Section 5).  

\section{Observations and Data analysis}
For the continuum data, an observing log
for both the GMRT and VLA observations as well as some of the
observed parameters of the continuum images are presented in Table~2                     
which is arranged as follows. Column 1: name of the telescope where we also list
the configuration for the VLA observations. In addition to our own data we have
also analysed many sets of archival VLA data on this galaxy. 
Column 2: the frequency of the observations;
columns 3 and 4: dates of the observations and
the time, t, spent on the source in minutes; 
columns 5 and 6: the phase calibrator used and its flux density estimated from the
observations; 
columns 7, 8 and 9: major and minor axes of the restoring beam
in arcsec and its position angle (PA) in degrees; column 10: the rms noise in
the continuum image in units of mJy/beam; columns 11 and 12: the peak brightness
and total flux density in units of
mJy/beam and mJy for the continuum images. The total error in the flux density is 
approximately 5\% for frequencies higher than 1 GHz and $\sim$10\% at 617 MHz.

\begin{table*}
  \caption{Observation log and observed parameters of the continuum images.} 
  \begin{tabular}{l r c r c c rrr rrr}
\hline
Telescope&Freq.&Obs.&t& Phase & S$_{\rm cal.}$&\multicolumn{3}{c}{Beam size} & rms & S$_{\rm pk}$ & S$_{\rm tot.}$ \\
     & MHz & date & min&Calib.&     Jy       & maj.    & min.      & PA      & mJy &   mJy        &   mJy          \\
    &      &      &   &       &  &$^{\prime\prime}$ & $^{\prime\prime}$ & $^\circ$ & /b &  /b &                 \\
 (1) & (2) & (3)  & (4)&(5)   &    (6)       & (7)     & (8)       & (9)     & (10) &   (11)      &   (12)     \\
 
\hline
GMRT &617   & 2002Jan11                 & 300 & 1120+057 & 3.57  & 46  &  23  &149 & 4       & 151  & 311   \\
VLA-A&1490       & 1986May24            & 236 & 1252+119 & 0.98  & 1.50& 1.50 &    & 0.04    & 19   &  87    \\
VLA-B&4860       & 1986Jul20            & 177 & 1252+119 & 0.62  & 1.50& 1.50 &    & 0.03    & 7.3  &  33    \\
VLA-A&4860       & 1986May24            & 44  & 1252+119 & 0.62  & 0.41& 0.36 &114 & 0.04    & 0.5  &  12    \\
VLA-A&8460       & 1998May13            & 23  & 1236+077 & 0.71  & 0.21& 0.19 &176 & 0.03    & 0.4  & 7.8   \\
VLA-AB&14939     & 1991Dec05            & 13  & 1252+119 & 0.52  & 0.57& 0.43 &127 & 0.19    & 1.0  & 8.9   \\
\hline  
\end{tabular}
\end{table*}

 \begin{figure*}
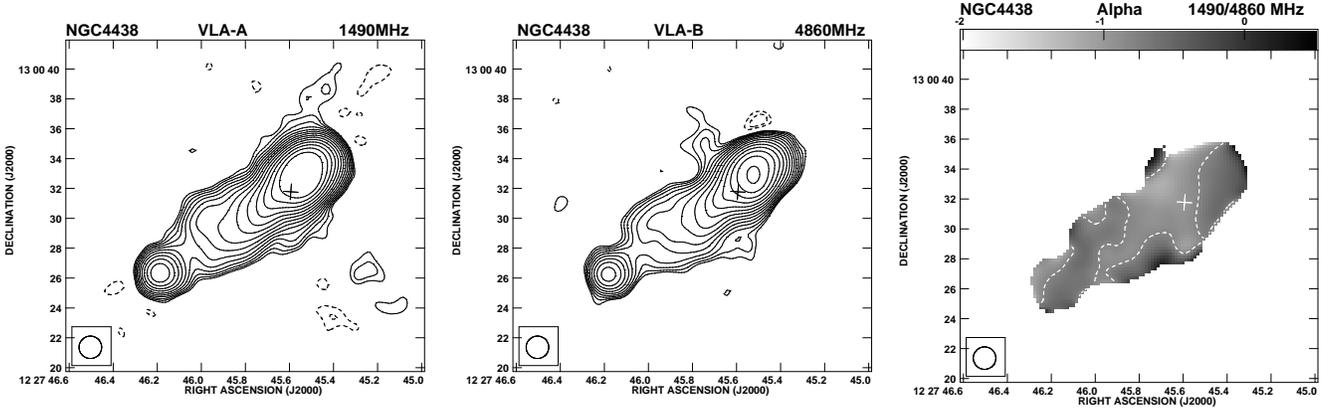

 \hbox{
   \psfig{file=N4438VALAP3R.PS,width=2.3in,angle=-90}
   \psfig{file=N4438VBCAP3R.PS,width=2.3in,angle=-90}
   \psfig{file=N4438SCL-CSP.PS,width=2.3in,angle=0}
    }
 \caption[]{Left panel: The VLA A-array image at 1490 MHz with an angular resolution
      of 1.5 arcsec. Contours: 0.037$\times$($-$4, $-$2.82, 2.82, 4, 5.65, 8 $\ldots$ ) 
      mJy/beam. 
           Middle panel: VLA B-array image at 4860 MHz with an angular resolution of 1.5 arcsec.
      Contours: 0.026$\times$($-$4, $-$2.82, 2.82, 4, 5.65, 8 $\ldots$ ) mJy/beam.
           Right panel: Spectral index image between 1490 and 4860 MHz. The contour is at $-$0.8.
      The $+$ sign marks the position of the radio nucleus in all the images.
               }
 \end{figure*}

 \begin{figure*}
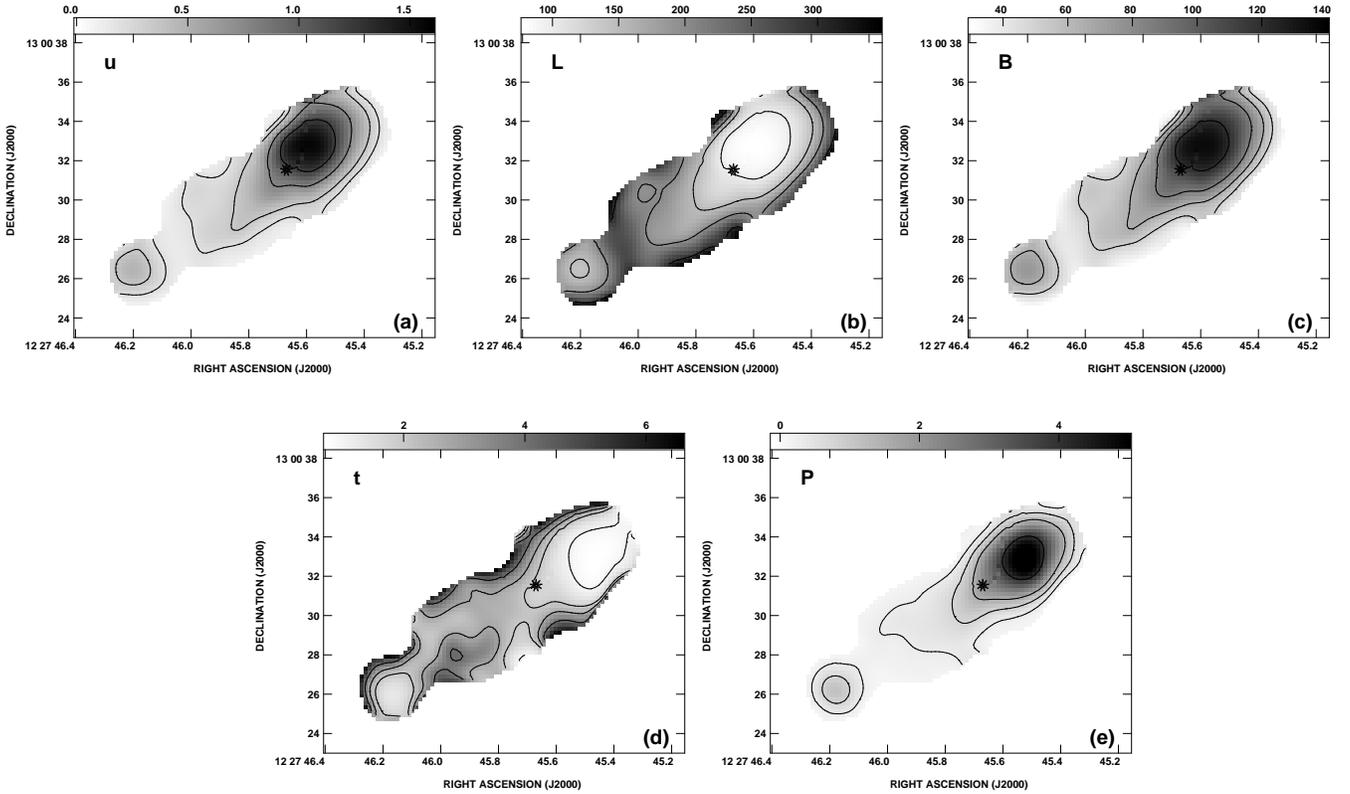

\vbox{
 \hbox{
   \psfig{file=U.PS,width=2.3in,angle=0}
   \psfig{file=L.PS,width=2.3in,angle=0}
   \psfig{file=B.PS,width=2.3in,angle=0}
    }
 \hspace{0.8cm}
 \hbox{
   \psfig{file=T.PS,width=2.3in,angle=0}
   \psfig{file=P.PS,width=2.3in,angle=0}
    }
      }
 \caption[]{
             Maps of the minimum energy parameters of the mini double-lobed radio source, made from
Fig.~2, assuming a heavy particle to electron energy ratio of k=40.  In each case, black denotes the
{\it highest} values and the star marks the position of the radio nucleus. (a)  The cosmic ray energy density, 
u$_{\rm CR}$.
Contours are $200$, $350$, $750$, and $1200$ eV cm$^{-3}$ and the peak value is $1.58\,\times\,10^3$ eV cm$^{-3}$ .
(b) Diffusion length, L$_{\rm D}$.  Contours are $100$, $150$, $200$ and $300$ pc  and the peak is $347$ pc.
(c) Magnetic field strength, B.  Contours are $50$, $65$, $85$, and $120$ $\mu$G and the peak is
$136$ $\mu$G. (d)  Lifetime of the particles, t.  Contours are $1$, $2$, $2.75$, and $4$ Myr and the
peak is $6.59$ Myr.  (e)  Cosmic ray power, P.  Contours are $0.33$, $0.75$, $1.5$, and $4$ $\times\,10^{39}$ 
ergs s$^{-1}$ and the peak is  $5.57\,\times\,10^{39}$ ergs s$^{-1}$.
               }
 \end{figure*}


\begin{figure*}
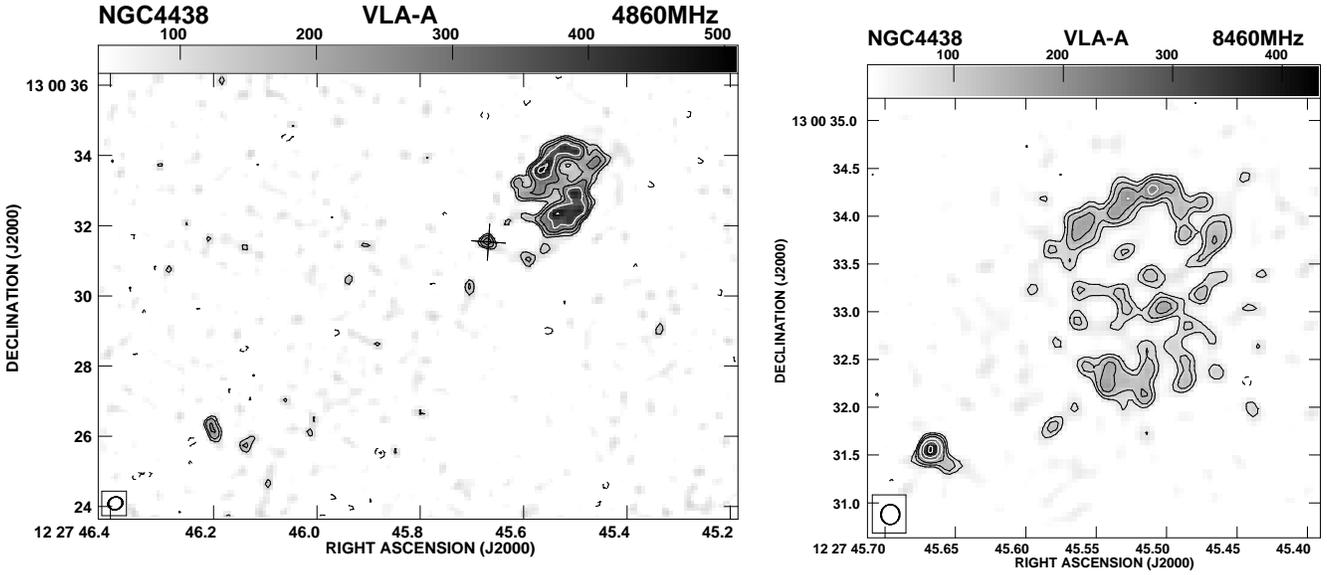

\hbox{
   \psfig{file=N4438VACAP2R.PS,width=4.0in,angle=-90}
    \psfig{file=N4438VAXP2.NU.PS,width=3.0in,angle=0}
   }
\caption[]{Left panel: The VLA A-array image at 4860 MHz with an angular resolution of $\sim$0.38 arcsec.
      Contours: 0.042$\times$($-$4, $-$2.82, 2.82, 4, 5.65, 8 $\ldots$ ) mJy/beam.
           Right panel: The VLA A-array image at 8460 MHz with an angular resolution of $\sim$0.19 arcsec
      showing only the nucleus and the western shell. 
      Contours: 0.025$\times$($-$4, $-$2.82, 2.82, 4, 5.65, 8 $\ldots$ ) mJy/beam.
           }
\end{figure*}


The radio continuum observations with the GMRT as well as with the VLA were made
in the standard fashion with each source observation interspersed
with observations of the phase calibrator.  The primary flux
density calibrator was 3C286 or 3C147 whose flux densities were estimated on the
Baars et al. (1977) scale using the 1999.2 VLA values. 
The bandwidth of the continuum observations with the GMRT at 617 MHz was 16 MHz,
while for the L-band observations it was 8 MHz. The bandwidth for all the
VLA continuum observations was 50 MHz.  The data analysis was done using the
Astronomical Image Processing System ({\tt AIPS}) of the National Radio
Astronomy Observatory (NRAO). Since GMRT data are acquired in the spectral-line mode 
with 128 spectral channels, gain and bandpass solutions were applied to each 
channel before combining them. Self-calibration was applied to all the GMRT and VLA 
data sets. The low-resolution 617-MHz map was made with the 3-D cleaning option in CLEAN  
using 16 fields. The self calibrated gains were applied to this data set 
correcting only for phase and rejecting all the failed solutions. 
For all the VLA data sets two cycles of phase and one cycle of amplitude 
self calibration were applied except for the VLA A-array 8460-MHz data where only two
cycles of phase self calibration were applied. 

The information for the H{\sc i} observations are presented later in Section~4 
(Table~5) where, in addition to the values described previously, 
we have also listed the spectral resolution in units of km s$^{-1}$ (column 6),
the rms noise in the channel maps in units of mJy/beam (column 10), the rms noise
in the spectrum in units of mJy (column 11) and the H{\sc i} flux density in
units of Jy km s$^{-1}$ (column 12).  The last row corresponds to the VLA-D array
data to which a broadscale taper has been applied (see Sect.~4.2).

The analysis of the \hbox{H\,{\sc i}} 
observations was also done in the standard way. 3C286 was the primary flux density and
bandpass calibrator. The total bandwidth for the GMRT \hbox{H\,{\sc i}} observations
was 8 MHz and the spectral resolution was 62.5 kHz. 
The total bandwidth for the VLA
D-array \hbox{H\,{\sc i}} observations was 6 MHz and the spectral resolution was 97.7 kHz.
Continuum subtraction was done by specifying line-free channels and using the AIPS task UVLIN.
The GMRT data were cleaned using the self-calibrated gains from the continuum data analysis.
For the VLA data analysis, the bright continuum source M87 was subtracted using UVSUB,
before the multi-channel data was continuum subtracted and CLEANed using IMAGR.

\section{Radio continuum emission}
\subsection{A mini double-lobed radio source}

The VLA A-array image at 1490 MHz
and the VLA B-array image at 4860 MHz with an angular resolution of
1.5 arcsec (Fig. 2: left and middle panels) show the well-known double-lobed
structure of the radio source in the nuclear region (Hummel \& Saikia 1991).
The double-lobed source has a total extent of $\sim$12 arcsec (960 pc), with the
western lobe being separated from the nucleus (the latter discussed in Section 3.2) by 
about 2.9 arcsec (230 pc) compared with 8.9 arcsec (730 pc) for the eastern lobe.
The spectral index image made from these scaled-array VLA images
by considering only those pixels which are 5 times above the rms noise
is also shown in Fig. 2 (right panel).
The spectral index, defined as S$\propto\nu^{\alpha}$, varies from $-$1.5 to $+$0.5, 
while the error in the spectral index
is $\sigma_\alpha\,=\,0.08$. The mean value of spectral index for the western lobe, 
the central region and
the eastern lobe are $-$0.78, $-$0.84 and $-$0.61 respectively, the spectral index
being steepest between the lobes.  

Given the steepness of $\alpha$, the emission
is clearly dominated strongly by synchrotron emission.  
However, it is important to quantify what, if any, contribution a thermal
component might make.  The latter would
flatten the spectral index from a steeper value, $\alpha_{NT}$, to the
flatter value, $\alpha$, that is observed.  In particular, we ask what
the thermal contribution would have to be in order to flatten the spectral
index by an amount that is significantly outside of our error bar, $\sigma_\alpha$,
in the spectral index map.
We use the measured flux densities of the two maps shown in Fig.~2, cut off to the
same $5\sigma$ level as the spectral index map.  These values are
$S_{1.5\, GHz}\,=\,86.3\,$mJy and $S_{4.86\, GHz}\,=\,\,32.5$ mJy, giving an
observed total spectral index of $\alpha\,=\,-0.825$.  If the true
non-thermal spectral index is $\alpha_{NT}\,=\,-0.825\,-\,2\sigma_\alpha\,=\,
-0.985$, then (following Lee et al. 2001) we find that the thermal flux density 
required to alter the spectral index by this much is
 $S_{1.5\,GHz}(th)\,=\,8.2$ mJy.  This corresponds to a required average electron
density of $n_e\,=\,7.2$ cm$^{-3}$ over a region whose equivalent spherical radius 
is 700 pc.  While there may be small regions within the beam in which the density 
is higher than this (e.g. Kenney \& Yale 2002), 
it is very unlikely that the electron density is this high,
on average, over the observed region.  Indeed, Machacek et al. (2004)
find a mean density of $n_e\,\approx\,0.03$ cm$^{-3}$ for the hot gas component
over a roughly equivalent volume.  
We therefore conclude that the thermal contribution
to these maps are negligible and continue with the minimum energy calculations
under the assumption that $\alpha\,\approx\,\alpha_{NT}$. 

\subsubsection{The minimum energy parameters}
The VLA images, which have the same spatial resolution and similar
signal-to-noise (S/N) ratios, along with the spectral index map
(Fig.~2) can be used to compute the following minimum energy parameters
(see Pacholczyk 1970; Duric 1991):
the cosmic ray energy density u$_{\rm CR}$, the cosmic ray electron 
diffusion length
L$_{\rm D}$, the magnetic field strength B, and the particle lifetime t.
This calculation can be done on a pixel-by-pixel basis assuming a 
particular geometry, as previously described in Irwin \& Saikia (2003).

In the case of the double-lobed source, we use
a geometry in which the line-of-sight depth is taken to be
the measured average width of the mini double-lobed
source, i.e. $4.21^{\prime\prime}\,=\,344$ pc with unity filling factor.  The
 lower- and higher-frequency cut-offs of the spectrum are $\nu_1$=10$^7$
and $\nu_2$=10$^{11}$ Hz, respectively, and we adopt two different values for
 the ratio of heavy$-$particle to electron
energy, k=40, and k=100.  The results are shown in Fig.~3 (a through d).
In Fig.~3e, we also show a map of power,
P=U/t, where U is the cosmic ray energy density u$_{CR}$ integrated
along a line of sight and t is the particle lifetime. The result
is identically P=(1+k)L, where L is the observed luminosity at a point.
Thus the map of P closely resembles the map of total flux density but does
not match it exactly because the computation of L requires an integration
over frequency which is dependent on spectral index, and the spectral index
is different at different points in the map. The map of P represents the
rate at which cosmic rays must be accelerated in order to maintain
equilibrium. 

Averages over the maps shown in Fig.~3 are given in Table~3.  We have
found that changes in the choice of line of sight distance and upper frequency cutoff
make relatively small changes in the results in comparison to the choice of k
(e.g. Irwin \& Saikia 2003).  For the large range in adopted value of k, 
the results, overall, are within a factor of $\approx\,1.5$ of each other.
Note also, that this choice affects only the absolute scale of the maps 
and not the point-to-point variations.  Beck \& Krause (2005)
have recently proposed adopting a system whereby the particle number ratio, rather
than the energy ratio, is used in such calculations.  For the mean spectral
index of the mini double-lobed source (${\bar \alpha}\,\sim\,-0.8$) the magnetic field
values using their formalism are within a factor of $\sim$2 of ours 
(using the classical approach).  As indicated in the table, the total power in the 
cosmic ray component is a few times $10^{42}$  ergs s$^{-1}$
and the total energy in cosmic rays is $10^{56}$ ergs.

\begin{table*}
 \centering
 \begin{minipage}{140mm}
  \caption{Minimum energy parameters.$^a$}
  \begin{tabular}{@{}lcccccc@{}}
\hline
 Model$^b$        & $\bar u_{CR}$$^c$ & $\bar {L_D}$$^d$ & $\bar B$$^e$ 
& $\bar t$$^f$ & $U_{CR}$$^g$ & $P_{CR}$$^h$\\
                  &    (eV cm$^{-3}$) &  (pc)    &  ($\mu$G)    &  (Myr)  & ($10^{55}$ ergs) & 
($10^{42}$ ergs s$^{-1}$)\\
\hline
k=40  &   463             & 178 & 68.6 & 2.46 & 8.75 & 1.72 \\
k=100 &   775             & 137 & 88.8 & 1.67 & 14.6 & 4.24  \\
\hline
\end{tabular}\hfill\break
$a$ See Pacholczyk (1970), Duric (1991) or Irwin \& Saikia (2003). \\ 
$b$ A line of
sight distance of $344$ pc is adopted. k is the ratio of heavy particle to electron
energy. \hfill\break
$c$ Cosmic ray energy density, averaged over the source. The total cosmic ray plus
magnetic field energy density is 7/4 times these values.\hfill\break
$d$ Average diffusion length.\hfill\break
$e$ Average magnetic field strength.\hfill\break
$f$ Average particle lifetime. \hfill\break
$g$ Total cosmic ray energy integrated over the source volume.\hfill\break
$h$ Total power, $P_{CR}\,$=\,(1\,+\,k)\,$L$, where $L$ is the luminosity radiated
by the electron component (see text).
\end{minipage}
\end{table*}


The estimates of the magnetic field, which determine the radiative lifetime
of the particles (see below), are $\sim$70 $\mu$G for k=40. It is of interest to compare this
value with other galaxies which have radio bubbles or lobes powered by an AGN.
Two of the well-studied galaxies with such features are NC6764 which has a total
linear extent of 2.6 kpc (Hota \& Saikia 2006) and Circinus which has an extent
of 8.1 kpc (Elmouttie et al. 1998). The equipartition fields in these two galaxies
for k=40 are $\sim$25 and 45 $\mu$G respectively. The value for NGC~4438 which is
more compact is slightly higher.

Of particular interest is the lifetime map (Fig.~3d) since it helps 
to determine whether or not the particles need to be accelerated in situ.
We use lobe locations identified by the well-defined peaks in the power map
(Fig.~3e), giving locations for the western lobe
of RA = 12$^h$ 27$^m$ 45.$^s$52  Dec $+$13$^\circ$ 00$^{\prime}$ 32.$^{\prime\prime}$9, 
and for the eastern lobe of
RA = 12$^h$ 27$^m$ 46.$^s$18  Dec $+$13$^\circ$ 00$^{\prime}$ 26.$^{\prime\prime}$1.  
These positions correspond to projected separations from the nucleus
of $2.6^{\prime\prime}$ (214 pc) and
$9.2^{\prime\prime}$ (758 pc) for the western and eastern lobes respectively.
At these positions, the particle lifetimes are, for k\,=\,40,  $t\,=\,0.84$ Myr (west)
and  $t\,=\,1.3$ Myr (east) (somewhat lower if k\,=\,100).  Ignoring a possible
line-of-sight component to the velocity, the required particle velocity is
only $250$ km s$^{-1}$ and $570$ km s$^{-1}$ for the west and east lobes,
respectively, if the relativistic particles are supplied by the radio nucleus.
These are lower limits, given the possible line-of-sight component to
the velocity, but they are sufficiently lower than $c$ that in situ 
acceleration is not required, based on lifetime arguments.
It is worth noting that the spectral index is typically flatter, the magnetic field 
stronger, and the particle lifetimes shorter at locations near the peaks of the two 
lobes in comparison to regions between the nucleus and lobes. This behaviour is typical 
of classical double-lobed sources where the peaks of emission are identified with 
regions where energy from the AGN is deposited.


\subsubsection{The eastern and western lobes/shells}

The higher resolution VLA A-array images at 4860 and 8460 MHz,
 with angular resolutions of $\sim$0.38
and 0.19 arcsec, respectively, are shown in Fig.~4.  These reveal
 the prominent
shell-like structure in the western lobe, noted earlier by Hummel \& Saikia (1991),
and also regions of emission from the eastern lobe. 

The structure of
the western lobe, which is seen more clearly here than in the image of 
Hummel \& Saikia, shows two main ridges of emission on the northern and
southern sides with the peak of emission being on the northern ridge rather than
the edge of the bubble farthest from the nucleus (Section 3.2). The western shell is also seen clearly 
in H$\alpha$ and x-ray wavelengths (Kenney \& Yale 2002; Machacek et al. 
2004). 
The eastern lobe is also known to
exhibit a shell-like structure which is seen clearly at H$\alpha$ and x-ray
wavelengths (Kenney \& Yale 2002; Machacek et al. 2004) although, in our high-resolution 
radio images, emission is seen
only from the south-eastern and southern parts of the shell.
The eastern shell, with a radius of $\sim$0.5 arcsec (40pc),
is smaller than the western one whose radius is
1.0 arcsec (80pc).  

Although there is a close correspondence between the
shells seen at radio and H$\alpha$ and x-ray wavelengths there are also 
significant differences.  The eastern
shell appears relatively more complete at H$\alpha$ and at x-ray wavelengths
whereas, as already noted, we have detected radio emission only from parts of the shell.
Moreover,  in the western lobe, both H$\alpha$ and x-ray emission peak close to the 
nucleus which is rather weak at radio wavelengths.

\subsection{The radio nucleus}
VLA A-array images at 4860 and 8460 MHz (Fig.~4) show clearly a compact radio source.
The position of this source is RA 12$^h$ 27$^m$ 45.$^s$67, 
Dec  +13$^\circ$ 00$^{\prime}$ 31.$^{\prime\prime}$54 at 4860 MHz, which agrees well with 
the position of  RA 12$^h$ 27$^m$ 45.$^s$66,
Dec +13$^\circ$ 00$^{\prime}$ 31.$^{\prime\prime}$53 determined from the 8460-MHz image.
These positions also agree with those of the optical and infrared nuclei 
(Clements 1983; Keel \& Wehrle 1993; Falco et al. 1999). The J2000 position of the
infrared peak, which is likely to be least affected by extinction, 
is RA 12$^h$ 27$^m$ 45.$^s$67, 
Dec  +13$^\circ$ 00$^{\prime}$ 31.$^{\prime\prime}$54 (Keel \& Wehrle 1993) and this value is also
consistent with the measurements at optical wavelengths.

The peak flux densities of the compact radio source estimated from our two images are
0.29 mJy/beam  at 4860 MHz and  0.43 mJy/beam  at 8460 MHz 
which yields an inverted radio spectrum with a spectral index of $+$0.69. 
The spectrum could be even more inverted if there is any contamination of the flux density 
at 4860 MHz by any extended emission. The highly inverted spectrum of this compact component 
suggests it to be the nucleus of NGC~4438 (see Section 5.1). It would be interesting to re-observe 
it for any possible signs of variability. The spectrum in this region appears steep in the 
low-resolution images (Fig. 2) because it is dominated by the extended bridge emission. 
The 8460-MHz image (Fig. 4, right panel) shows evidence of a jet-like extension from the nucleus
along a PA of $\sim$233$^\circ$ which is almost orthogonal to the orientation of the mini-double. 
The nature of this feature is unclear; a more sensitive image to trace its extent would be useful.
  
\begin{figure}
\hbox{
  \psfig{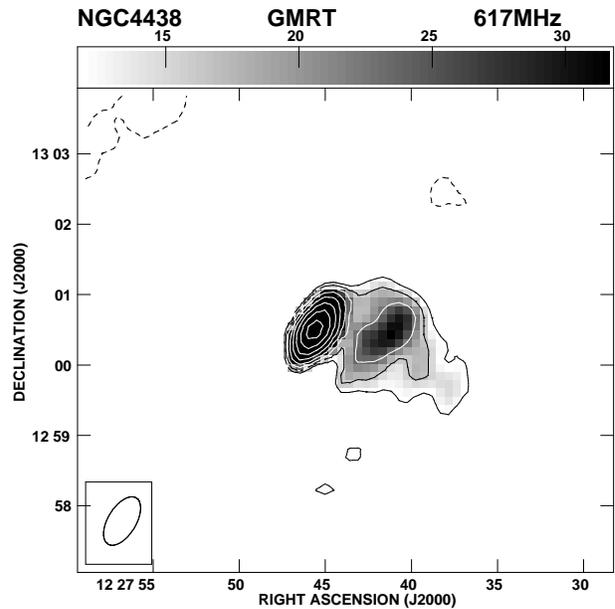}
   }
\caption[]{The GMRT image of NGC~4438 and 617 MHz with an angular resolution 
           of $\sim$33 arcsec.
          Contours: 4$\times$($-$4, $-$2.82, 2.82, 4, 5.65, 8 $\ldots$ ) mJy/beam.
          } 
\end{figure}

\subsection{The western extended radio emission}
 
In Fig.~5, we show the extended 617 MHz continuum emission from NGC~4438,  observed with the GMRT,
over a field of view $\sim$400 times larger in area than that shown in Fig.~2. 
The brightest emission, seen as an unresolved component at the centre
of the map, contains the nucleus and the mini-lobe discussed earlier (Fig.~2).  The western extended 
radio emission was initially reported by Allen et al. (1973) and Kotanyi, van Gorkom \& Ekers (1983).  
The peak of the extended emission is about $\sim$1 arcmin from the nuclear region and
extends over 2 arcmin from the nucleus of NGC~4438, or at least 10 kpc towards the south-west.

The rms noise in Fig.~5 is 4 mJy/beam, which is higher than desirable, due to the presence of the strong
radio source M87. This is also possibly responsible for the significant variations in the total
flux density values of NGC~4438 which have been reported in the literature (see Table 4). 
Vollmer, Thierbach \& Wielebinski (2004a) have tried to determine systematically the radio continuum
spectra of galaxies in the Virgo cluster and quote a  spectral index of $-$0.67 for the total
emission from NGC~4438 using a number of measurements between 600 and 10550 MHz.

\begin{table}
  \caption{Radio flux density}
  \begin{tabular}{c c c l}
\hline
 Frequency & S$_{\rm tot.}$ & S$_{\rm double}$ & Reference \\
 MHz       &   mJy          &    mJy         &           \\
 (1)       &    (2)         &   (3)          &   (4)      \\  
\hline
610        &  324$\pm$30    &                &  VTW2004; G2003   \\
617        &  311$\pm$30    &  151$^a$       &  P         \\ 
1400       &  150$\pm$10    &   86$^b$       &  KE1983    \\
1420       &  149$\pm$15    &                &  VTW2004; G2003   \\
1490       &                &   87           &  P; HS1991    \\
4850       &  97$\pm$9      &                &  NKW1995; VTW2004   \\
4850       &  70$\pm$10     &                &  BWE1991    \\
4860       &                &   33           &  P; HS1991    \\
8600       &  49$\pm$4      &                &  VTW2004       \\
10550      &  44$\pm$4      &                &  NKW1995; VTW2004  \\
\hline
\end{tabular}

$^a$ Estimated from the peak flux density in our GMRT image. \\
$^b$ Nuclear flux density from KE1983, which corresponds to the 
     inner double. \\

References. BWE1991: Becker et al. 1991; G2003: Gavazzi et al. 2003; 
HS1991: Hummel \& Saikia 1991; 
KE1983: Kotanyi \& Ekers 1983; NKW1995: Niklas et al. 1995;
P: Present paper; VTW2004: Vollmer et al. 2004a.

\end{table}

\begin{figure}
\hbox{
  \psfig{file=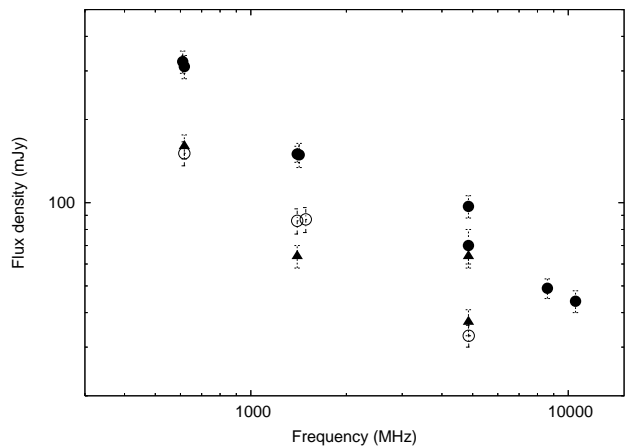,width=3.4in,angle=-90}
   }
\caption[]{The continuum flux densities of the total emission (filled circles), the 
           inner double (open circles) and the western extended emission (filled triangles)
           of NGC4438 using the flux densities listed in Table 4. 
          }
\end{figure}


Subtracting the flux densities of the nuclear double-lobed
source for which we have reliable measurements of the total flux density (Table 4)
from the measurements of total flux density we can estimate the spectral index
of the western extended emission.
Using the measurements of Vollmer et al. for the total flux density
gives flux densities of 62 and 64 mJy for the western extended emission
at 1400 and $\sim$5000 MHz respectively.
This yields a flat spectral index of $\sim$0 for the 
extended emission, even flatter than the value of $\sim$$-$0.5 reported by Kotanyi et al. (1983).
Adopting the value of 70 mJy for the total flux density at 5000 MHz (Becker et al. 1991) gives 
us a flux density of 37 mJy and a spectral index of $\sim$$-$0.4 for the extended emission. The
western extended radio emission is visible in the 10550 MHz image of 
Niklas, Klein \& Wielebinski (1995) possibly due to its flat spectrum.
Estimating the flux density of the nuclear emission at 10550 MHz by extrapolating it from lower
frequencies and subtracting it from the total flux density also yields a similar flat spectral
index of $\sim$$-$0.4 between 1400 and 10550 MHz.

At low-frequencies the total flux density at 610 MHz (Vollmer et al. 2004a)  is consistent
with our estimate of 311$\pm$30 mJy at 617 MHz with an angular resolution of
46$^{\prime\prime}$$\times$24$^{\prime\prime}$ arcsec$^2$ along a PA of 149$^\circ$.
Identifying the peak flux density of 151 mJy in the GMRT
image at 617 MHz with the nuclear double gives a flux density of $\sim$160 mJy for the extended emission.
The peak flux density of 151 mJy is consistent with the
extrapolated flux density of $\sim$179 mJy at 617 MHz for the nuclear double using our VLA measurements.
This yields a spectral index of $\sim$$-$1.1 for the extended emission between 600 and 1400 MHz significantly
steeper than the high-frequency spectrum (see Fig. 6). This suggests that the low-frequency spectral index is dominated
by non-thermal emission, while at high-frequencies contributions from thermal emission become important.
This is a demonstration of both thermal and non-thermal radio-emitting material exisiting in
the extra-planar gas. It is relevant to note that a flat spectral index of $\sim$$-$0.5 for extended
extra-planar radio emission may also be due to re-acceleration of particles (e.g. Bell 1978;
Duric 1986) as has been suggested for the Virgo cluster galaxy NGC~4522 (Vollmer et al. 2004b). 
However, the difference in spectral index discussed above suggests that this is not 
the explanation for NGC~4438.

\begin{figure}
\hbox{
  \psfig{file=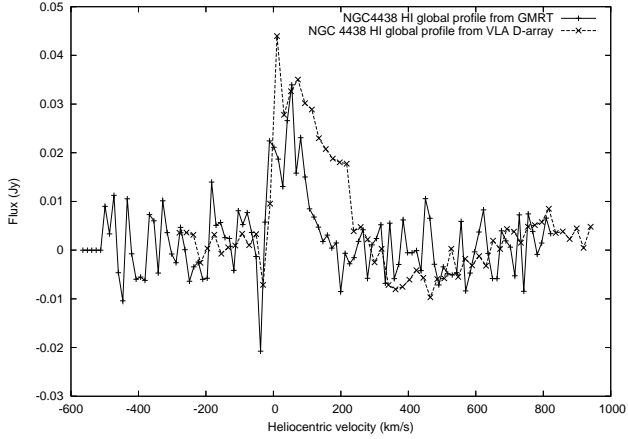,width=3.3in,angle=-90}
   }
\caption[]{The global profile of H{\sc i} emission from the GMRT observations with
an angular resolution of $\sim$36 arcsec (continuous line) and VLA D-array observations
tapered to an angular resolution of  $\sim$125 arcsec (dashed line).   
          }  
\end{figure}

\begin{figure}
\hbox{
  \psfig{file=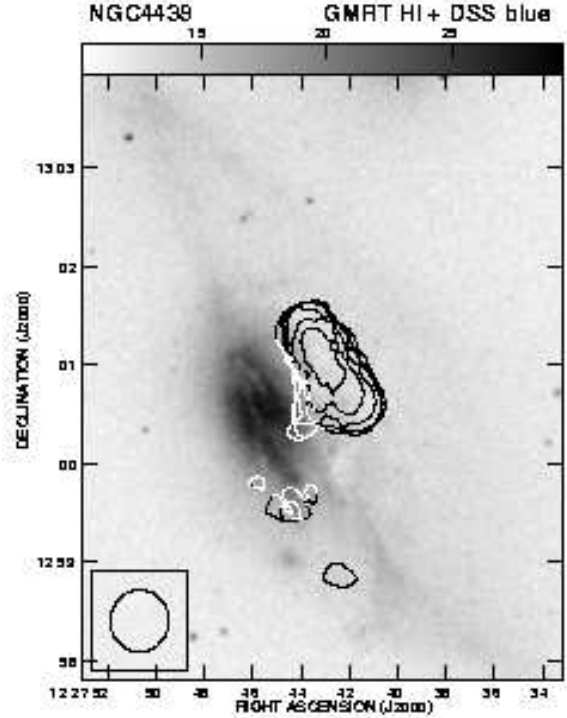,width=3.0in,angle=0}
   }
\caption[]{The moment zero image from the GMRT observations with an angular resolution 
           of $\sim$36 arcsec superimposed on the DSS blue-band image.
           Contour levels are (11.28, 16, 22.60, 32 $\ldots$)$\times$10$^{19}$ atoms cm$^{-2}$, 
           in steps of $\sqrt{2}$.
           }
\end{figure}

 
\section{H{\sc i} emission}

\begin{table*}
  \caption{Observation log and observed parameters of the H{\sc i}-images.}
  \begin{tabular}{lccccr rrr rrc}
\hline
 Tel. & Obs.  & t &Phase & S$_{\rm cal}$ & Vel & \multicolumn{3}{c}{Beam size} &   map    & spec.   & H{\sc i}  \\
      & date  &min&calib.&     Jy        & res.& maj.    & min.      & PA             &  rms           &  rms  &  flux    \\
    &    &   &      &        &km/s & $^{\prime\prime}$ & $^{\prime\prime}$ & $^\circ$ & mJy/b & mJy     &Jy km/s     \\
(1) & (2)&(3) & (4) & (5)    &(6) &                (7) &              (8) &      (9) &  (10) &     (11) &    (12)   \\ 
\hline
GMRT  & 2002Apr14 & 300 & 1254+116 & 0.79 &13.2       & 38       &35         &178            & 1.2         &5.8         & 2.9        \\
VLA-D & 1988Jul02 & 330 & 1252+119 & 0.92 & 20.7      &  58      & 48        & 0             & 0.9         &2.0         & 4.0     \\
      &           &     &          &      &           & 127      &124        &84             & 1.2         &4.6         & 6.5        \\
\hline
\end{tabular}\hfill\break
\end{table*}

H{\sc i} observations of NGC~4438 have been reported earlier with an angular resolution
of 23$\times$118 arcsec$^2$ along PA=0$^\circ$ using the Westerbork telescope by Kotanyi (1981),
and more recently with an angular
resolution of $\sim$40 arcsec (3.3 kpc) using the VLA-D array by Cayatte et al.
(1994) and again with the VLA D-array by Li \& van Gorkom (2001).
We observed this source with the GMRT with the goal of
determining the structure of H{\sc i} in emission
with higher resolution, and also to detect H{\sc i} in absorption
towards the central compact source to identify any kinematic effects of the nuclear activity
on the H{\sc i} gas. Our highest resolution (see Table~5) was slightly higher than
obtained in previous observations.  To supplement our GMRT observations, we have
also analysed VLA D-array archival data to detect and study the
disk emission, which has not been seen earlier, with high surface brightness sensitivity.
We have also tapered the data to create another data set which emphasizes broadscale structure
in order to detect and understand the properties of more extended H{\sc i} gas. 
The phase centre of this archival data set was the centre of NGC~4438, unlike the observations
reported by Li \& van Gorkom (2001). 
The VLA results presented here are of higher surface brightness
sensitivity than those reported earlier and show several new features which we discuss
in this paper. 

\begin{figure*}
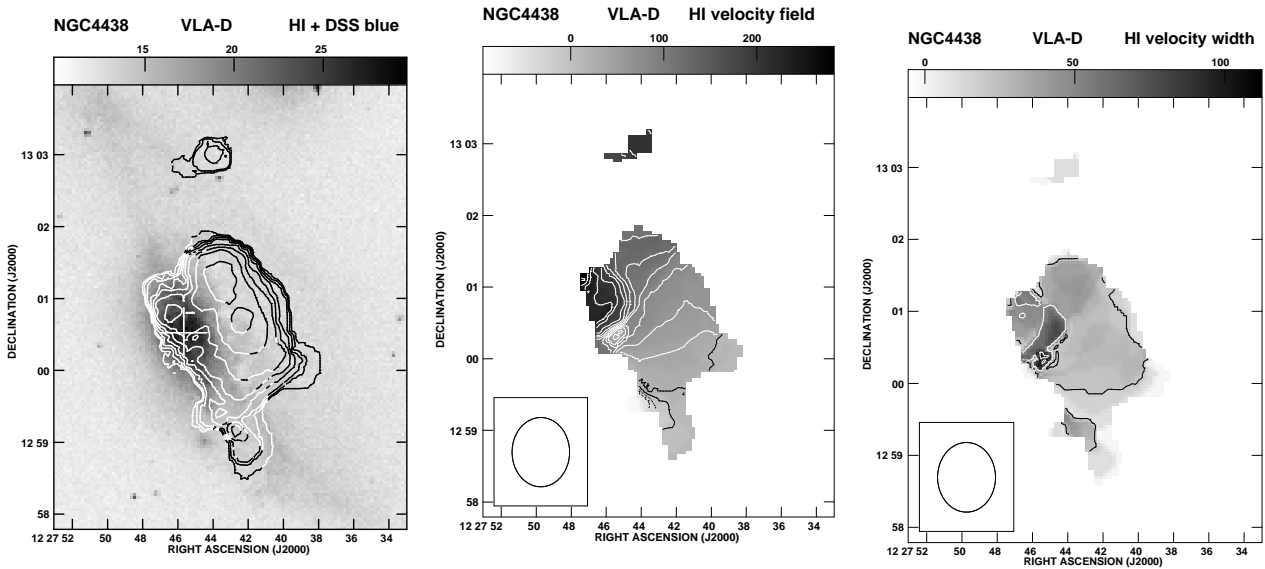

\hbox{
  \psfig{file=4438VDBMM0DBL.NEW.PS,width=2.2in,angle=0}
  \psfig{file=4438VDBMM1.ZM.PS,width=2.2in,angle=0}
  \psfig{file=4438VDBMM2.ZM.PS,width=2.2in,angle=0}
   }
\caption[]{Left panel: The moment zero image from the VLA D-array observations with an 
           angular resolution of $\sim$53 arcsec superimposed on the DSS blue-band image.
           Contour levels: (3.84, 5.45, 7.69 $\ldots$)$\times$10$^{19}$ atoms cm$^{-2}$ in steps of $\sqrt{2}$.
           Middle panel: The corresponding moment one image showing the velocity field.
           Contour levels from south to north are $-$40, $-$20, 0, 20, 40, 60, 80 $\ldots$ km s$^{-1}$ in steps of 
           20 km s$^{-1}$.
           Right panel: The corresponding moment two map showing the velocity dispersion.
           Contour levels: 20, 50 and 100 km s$^{-1}$.
           }
\end{figure*}
\begin{figure*}
\hbox{
  \psfig{file=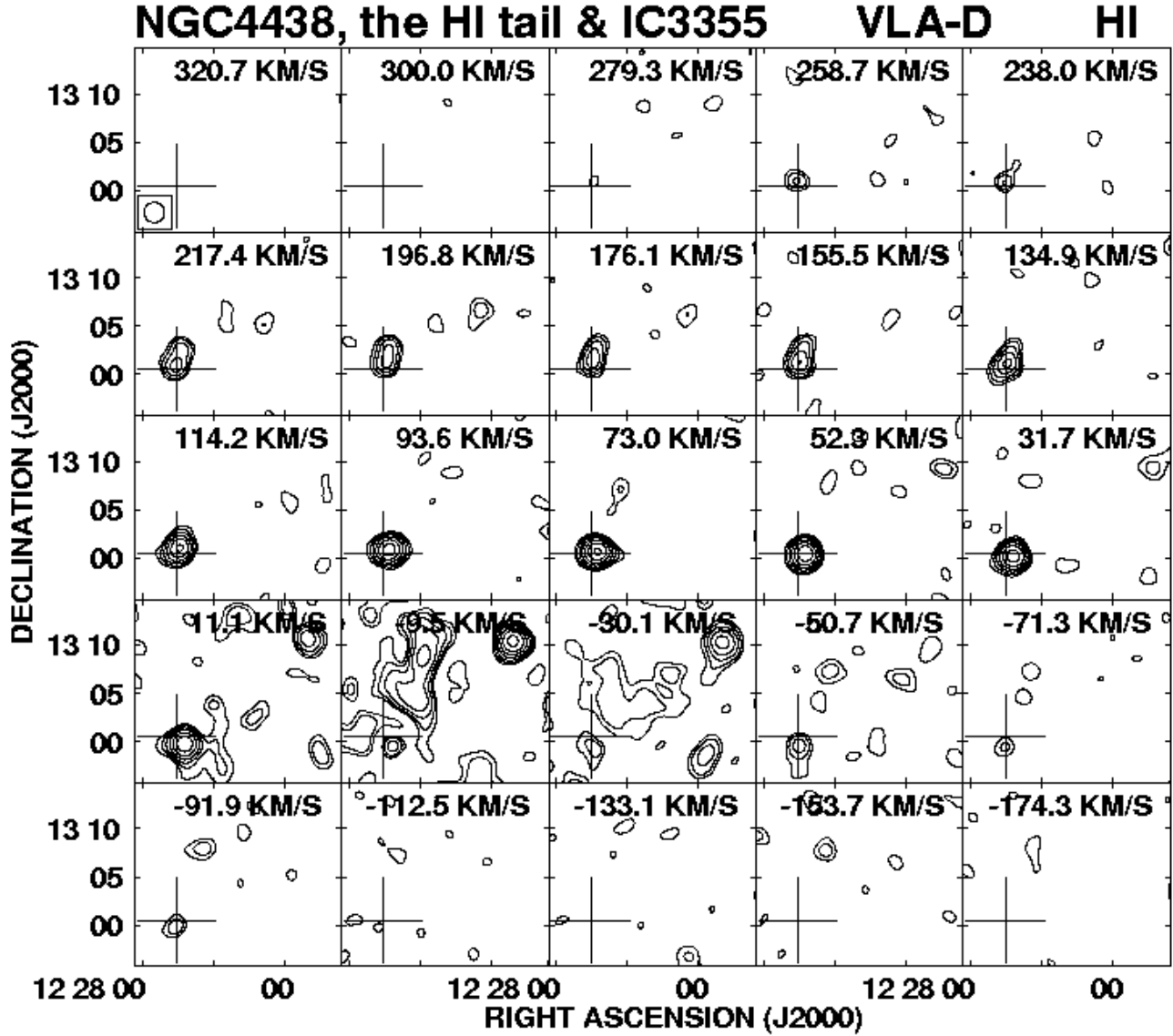,width=7.0in,angle=-90}

   }
\caption[]{ H{\sc i}-emission channel maps made with a resolution of $\sim$125 arcsec.
            The contours levels are 1.2 $\times$ ($-$4, $-$2.82, 2.82, 4, 5.65 $\ldots$) mJy/beam.
            The systemic velocity of NGC~4438 is  71 km s$^{-1}$.
            The velocity separation between adjacent channels is 20.7 km s$^{-1}$.
            The crosses in all the images denote the position of the radio nucleus of NGC~4438.
The irregular galaxy, IC~3355, is located at
RA 12$^h$ 26$^m$ 51.$^s$13, Dec  +13$^\circ$ 10$^{\prime}$ 32.$^{\prime\prime}$6 (see Sect.4.2.1).
          }
\end{figure*}

\begin{figure*}
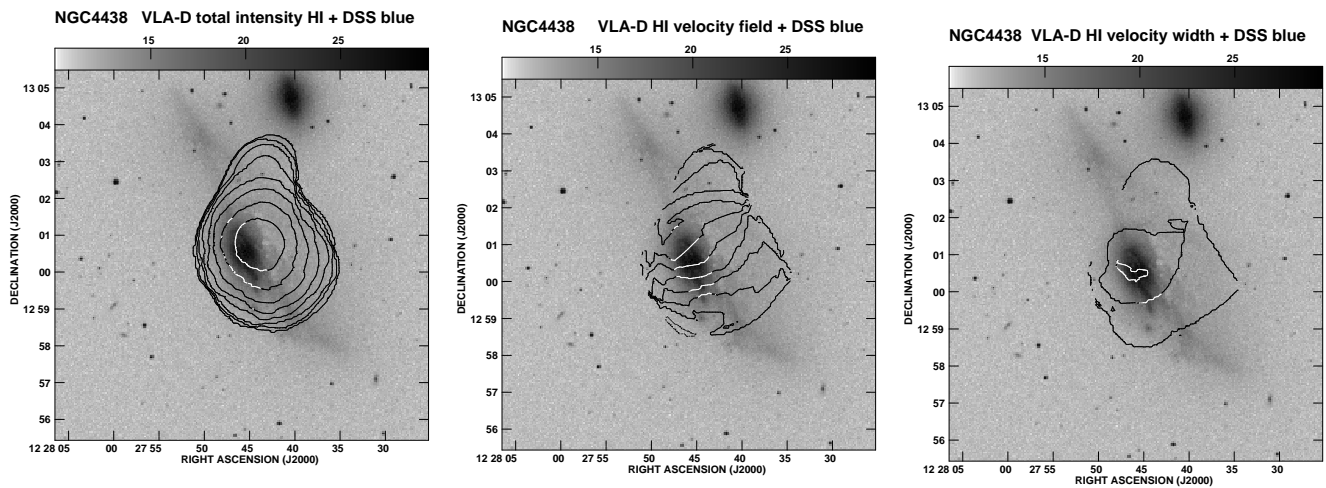

\hbox{
  \psfig{file=N4438DSSBL-VDHIM0.PS,width=2.3in,angle=0}
  \psfig{file=N4438DSSBL-VDHIM1.PS,width=2.3in,angle=0}
  \psfig{file=N4438DSSBL-VDHIM2.PS,width=2.3in,angle=0}
   }
\caption[]{Left panel: The moment zero image from the VLA D-array observations tapered to an
           angular resolution of $\sim$125 arcsec superimposed on the DSS blue-band image.
           Contour levels: (2.26, 3.22, 4.54, $\ldots$)$\times$10$^{19}$ atoms cm$^{-2}$ in steps of $\sqrt{2}$.
           Middle panel: The corresponding moment one image showing the velocity field.
           Contour levels from south to north are $-$5, 20, 40, 60, 80, 100, 120, 140, 160, 180 and 190 km s$^{-1}$. 
           Right panel: The corresponding moment two map showing the velocity dispersion.
           Contour levels are 20, 55 and 70 km s$^{-1}$.
          }
\end{figure*}


\subsection{GMRT results}

We present here the global profile (Fig. 7) and the
total-intensity H{\sc i} image from the GMRT observations (Fig. 8).
These observations clearly show an elongated structure along a PA of $\sim$35$^\circ$
which is very similar to the PA of 29$^\circ$ for the central stellar disk measured from
molecular line observations (Kenney et al. 1995). This elongated structure extends for $\sim$6.5 kpc
along the major axis in our GMRT image. 
It is roughly parallel to the stellar disk and $\sim$4.1 kpc away from its mid plane.
Earlier H{\sc i} images (e.g. Kotanyi et al. 1983; Cayette et al. 1994;
Li \& van Gorkom 2001) show the displacement of H{\sc i} gas on the western side of the
stellar disk but these observations show the elongated structure more clearly.
In addition to this elongated structure there are weak blobs of H{\sc i} emission towards the south.

The H{\sc i} mass of the elongated feature is 1.5$\times$10$^8$M$_\odot$, while
the total \hbox{H\,{\sc i}} mass estimated from the global profile (Fig. 7) obtained from the
GMRT observation is $\sim$2$\times$10$^8$M$_\odot$. The global profile exhibits a sharp
drop in intensity at $\sim$$-$40 km s$^{-1}$ and a tail of emission on the red-shifted side.
Although most of the \hbox{H\,{\sc i}} gas appears red-shifted with respect to the systemic
velocity of the galaxy (71 km s$^{-1}$), the blobs towards
the south appear blue shifted. The velocity gradient of the elongated structure shows
that the north-eastern side is receding with a maximum heliocentric radial velocity
of  107 km s$^{-1}$ while the south-western side is approaching with a minimum radial velocity
of $-$28 km s$^{-1}$. These velocities are consistent with the Westerbork observations reported
by Kotanyi (1981). The middle point of the radial velocity range
of the elongated feature from our observations is $\sim$40 km s$^{-1}$.
The maximum line width in this elongated H{\sc i} emission structure is $\sim$40 km s$^{-1}$.
The observed sense of rotation of this feature is similar to that of CO emission seen in
the stellar disk of the galaxy (Combes et al. 1988; Kenney et al. 1995; Vollmer et al. 2005), and
lies close to the position of the extra-planar CO(1-0) emission along a PA$\sim$20$^\circ$
(Vollmer et al. 2005).  The extent of the
extra-planar CO emission is very similar to that of the H{\sc i} feature seen here,
although the peak of the CO emission lies slightly to the north-west while that of the
H{\sc i}-elongated structure towards the centre of the feature. In this region the DSS blue
band image shows the presence of dust plumes which appear very similar in nature to those
seen in the Virgo cluster galaxy NGC~4402 (Crowl et al. 2005).

\subsection{VLA results}
We first present the results of the VLA D-array full-resolution observations with an angular resolution
of 58$\times$48 arcsec$^2$ along a PA of 0$^\circ$. Moment maps were made with a cutoff at
4$\sigma$ and integrated in velocity from $-$133 to velocity $+$340 km s$^{-1}$ (Fig. 9). With a higher surface
brightness sensitivity than the GMRT observations, the total-intensity contours of H{\sc i} emission show
that the elongated feature extends further south with a total size of $\sim$2$^\prime$ (9.8 kpc) along a similar
orientation to that of the elongated feature seen in the GMRT observations. There is also H{\sc i} emission from
the optical disk of the galaxy with the emission having a sharper edge towards the north and extending further
towards the south with a similar total extent $\sim$2$^\prime$ (9.8 kpc). There is a clump of H{\sc i} emission at
the southern end of the disk emission with velocities in the range of $\sim$$-$20 to $+$20 km s$^{-1}$.
The moment 0 map also shows a clump of H{\sc i} emission towards the north between NGC~4438 and NGC~4435
with heliocentric velocities ranging from $+$176 to $+$217 km s$^{-1}$.

The velocity field of NGC~4438 in this VLA D-array map is complex.
In the central region of the disk within $\sim$10 arcsec of the nucleus the isovelocity contours are
orthogonal to the central stellar disk with velocities ranging from 40 km s$^{-1}$ on the southern side
to 240 km s$^{-1}$ on the northern side, yielding a gradient of 10 km s$^{-1}$ arcsec$^{-1}$.
At larger distances from the nucleus the isovelocity contours along the disk of the galaxy
range from $-$80 near the southern clump to 280 km s$^{-1}$ on the northern edge giving a velocity
gradient of 3 km s$^{-1}$ arcsec$^{-1}$. There is a hint of the isovelocity contours exhibiting a 
C-shaped curve or its reflection towards the north-east and south-west regions of the extra-planar gas,
which requires confirmation from observations
of higher spatial resolution. If confirmed, this would be reminescent of `backward question marks' shaped
isovelocity contours in NGC~4522 (Kenney, van Gorkom \& Vollmer 2004) which is at a similar
orientation and undergoing interaction with the ICM of the Virgo cluster. The extra-planar gas in NGC~4438 has
a smoother velocity field than the disk ranging from $+$10 on the southern side 
to 130 km s$^{-1}$ on the northern side
giving a lower velocity gradient of 1 km s$^{-1}$ arcsec$^{-1}$ over a similar length scale.
It is worth noting that the blue-shifted velocities of $-$28 km s$^{-1}$ in the southern side of the
elongated feature seen in  the GMRT and Westerbork observations is not apparent in the D-array image,
suggesting that observations with different resolutions pick up sub structures with different velocities.

Approximately 15 arcsec south of the nucleus centred at RA $\sim$12$^h$ 27$^m$ 45.$^s$4, 
Dec 13$^\circ$ 00$^\prime$ 20$^{\prime\prime}$, there is a
small region of 15$\times$10 arcsec$^2$ along a PA$\sim$130$^\circ$ where the iso-velocity contours are closed
and have values ranging from 40 to 80 km s$^{-1}$. These contours represent gas which is approaching us
relative to the gas in its vicinity. The moment 2 map shows that the velocity
dispersion in this region is high with values in the range of 90$-$115 km s$^{-1}$.
The moment 2 map also shows that towards the north of this region, the velocity dispersion is typically
50$-$80 km s$^{-1}$ over a total extent of $\sim$3.7 kpc approximately orthogonal to the PA of the disk.
For the rest of the galaxy and the extra-planar emission the line widths are in the range of 20$-$50
km s$^{-1}$ with the line widths being larger on the northern side.

The VLA D-array data were also imaged with a 1k$\lambda$ taper (resolution of $\sim$ 125 arcsec) which
shows more H{\sc i} gas primarily on the western side. The global
profile (Fig. 7) superimposed on the one from the GMRT higher resolution observations shows that there
is indeed more H{\sc i} gas towards the red-shifted side. The VLA spectrum also shows a
sharp cut off on the blueward side of the spectrum as seen in the GMRT global profile.
The total H{\sc i} flux density is 6.5 Jy km s$^{-1}$ which is consistent with that of Arecibo
mesurements by Giovanardi, Krumm \& Salpeter (1983) who quote a line flux density of
6.1$\pm$0.6 Jy km s$^{-1}$.
Our VLA mesurements correspond to a total H{\sc i} mass of 4.4$\times$10$^8$ M$_\odot$
which is $\sim$2 times larger than that estimated from the GMRT spectrum.

The channel maps obtained with a spatial resolution of $\sim$ 125 arcsec are shown in Fig  10.
In addition to the H{\sc i} emission associated with NGC~4438 whose nucleus is 
marked with a cross in the figure,  
H{\sc i} emission is also seen at the position of the irregular
galaxy, IC~3355 (see figure caption), and a large cloud of 
H{\sc i}, which we will refer to as the `tail', is
detected which is most prominent in the velocity channel at $\sim$$-$9.5 km s$^{-1}$.
These features will be discussed more fully in the next two sections.

Moment maps of NGC~4438 were made of the tapered VLA data
with a cutoff at 4$\sigma$ and integrated in velocity from
$-$71 to 238 km s$^{-1}$ (Fig. 11).  The H{\sc i} total-intensity image with an angular
resolution of 127$\times$124 arcsec$^2$ looks resolved with an extension towards the north. The
northern clump seen in the full-resolution D-array image is coincident with this extension.
In the tapered image the disk emission, the extra-planar elongated structure and the southern
and northern clumps appear blended with more diffuse H{\sc i} gas.

The velocity contours in the southern side of the disk are approximately
parallel to each other along a PA of $\sim$$-$80$^\circ$ while those on the northern side are oriented along
a PA of $\sim$$-$45$^\circ$. The velocity in the disk ranges
from $+$20 to 140 km s$^{-1}$ with a velocity gradient of 0.9 km s$^{-1}$ arcsec$^{-1}$. The
extra-planar gas, including the northern extension, has velocities ranging from $\sim$20 to
$\sim$190 km s$^{-1}$ yielding an average velocity gradient
of $\sim$0.5 km s$^{-1}$ arcsec$^{-1}$.  At this resolution the H{\sc i} gas in the disk as well
as the extended emission on the western side are rotating about a heliocentric velocity of
$\sim$110 km s$^{-1}$ about an axis which is at a PA of $-$48$^\circ$. 
There is some hint that the isovelocity contours of the extra-planar gas seen
in this tapered image appear to curve inwards about this axis at larger distances from the disk, 
while in the full-resolution image they may diverge closer to the interface between the disk and the
extra-planar gas, reminescent of the `backward question mark' in  
the Virgo cluster galaxy, NGC~4522, which has been affected by ram-pressure stripping 
(Kenney, van Gorkom \& Vollmer 2004). 
Some evidence of similar diverging iso-velocity contours can be seen in the galaxy
NGC~2820, which is affected by ram pressure stripping due to the intra-group medium (Kantharia et al. 2005).
It may be relevant to study the velocity structures in the wakes of ram-pressure stripped gas with
features similar to a von-Karman vortex street (e.g. Fig. 8 of Roediger, Br\"uggen \& Hoeft 2006) to
understand the above-mentioned iso-velocity contours.    
The moment 2 map shows that the width of the H{\sc i} line changes from $\sim$20 km s$^{-1}$
in the outer regions to $\sim$87 km s$^{-1}$ near the centre of the galaxy.

It is also of interest to compare the H{\sc i} velocity field with those obtained at other
wavelengths, although one must bear in mind that the H{\sc i} observations are of much coarser
resolution than at other wavebands.  The velocities of 
CO and H$\alpha$ emission along PAs of 29$^\circ$ and 27$^\circ$
respectively, and H$\alpha$ gas along a PA of 29$^\circ$, all passing through
the nucleus of the galaxy, have been presented by Kenney et al. (1995) and Chemin et al. 
(2005) respectively. The optical observations having a seeing of 1.5$-$2 arcsec, the CO observations
having an angular resolution of $\sim$6 arcsec while the H{\sc i} images have angular resolutions
of $\sim$50 and 120 arcsec for the full-resolution and tapered images respectively.
As noted by Kenney et al. the CO and H$\alpha$ velocities are symmetric and in good agreement
within a galactocentric radius of $\sim$10 arcsec. Beyond this distance, the velocities become
asymmetric with the CO velocities on the north-eastern side increasing to $\sim$240 km s$^{-1}$ at
$\sim$20 arcsec from the nucleus while the H$\alpha$ velocities increase to $\sim$275 km s$^{-1}$
at $\sim$40 arcsec from the nucleus. On the south-western side the CO extends to $\sim$15 arcsec
with a minimum blue-shifted velocity of $-$105 km s$^{-1}$ but the H$\alpha$ gas seen by Kenney et al.
extends up to $\sim$130 arcsec with a nearly constant velocity of $\sim$$-$20 km s$^{-1}$.
The velocities obtained by Chemin et al. (2005) are very similar to those of Kenney et al. on the
north-eastern side, but goes to significantly higher blue-shifts of $\sim$$-$150 km s$^{-1}$
approximately 15 arcsec south-west of the nucleus.  
The H{\sc i} velocity field with a full-resolution
of $\sim$53 arcsec (Fig. 9) exhibits a similar pattern with velocities ranging from $\sim$$-$50 km s$^{-1}$
on the south-western side to $\sim$260 km s$^{-1}$ on the north-eastern side
$\sim$70 arcsec from the nucleus. 


\subsubsection {IC~3355}

IC~3355 (VV 511, DDO 124, VCC 0945) is an irregular galaxy (Fig. 12)
located $\sim$82 kpc from NGC4438.  This galaxy is listed
in the atlas of interacting galaxies by Vorontsov-Vel'Yaminov (1977)  
who also note the presence in its optical tail of `3 diffuse companions in a
blue haze'. The B-band image shows a number of compact star-forming regions towards the
eastern side of the galaxy. Spectroscopic observations at both radio and optical wavelengths
listed in NED, the HyperLeda (http://leda.univ-lyon1.fr; Paturel et al. 2003) and 
GOLDmine (http://goldmine.mib.infn.it; Gavazzi et al. 2003) data bases and available 
information in the literature show the presence of two systems, one with a heliocentric 
velocity of $\sim$$-$10 km s$^{-1}$ and the other at a heliocentric velocity of 
$\sim$162 km s$^{-1}$ (de Vaucouleurs, de Vaucouleurs \& Nieto 1979; Sulentic 1980;
Haynes \& Giovanelli 1986; Hoffman et al. 1987; Cayatte et al. 1990; 
Huchra, Geller \& Corwin 1995; Gavazzi et al. 2004). 
At radio wavelengths Haynes \& Giovanelli (1986) report the
detection of a weak feature at 162 km s$^{-1}$ and note that the `spectrum in severely
blended with local H{\sc i} emission'. Our spectrum does not show any feature
at $\sim$162 km s$^{-1}$ (Fig. 12), and this velocity is perhaps incorrect. 
Hoffman et al. (1987) and Cayatte et al. (1990) 
have reported the detection of H{\sc i} emission at a heliocentric velocity of 
$-$9 and $-$12 km s$^{-1}$ respectively.

\begin{figure}
\vbox{
  \psfig{file=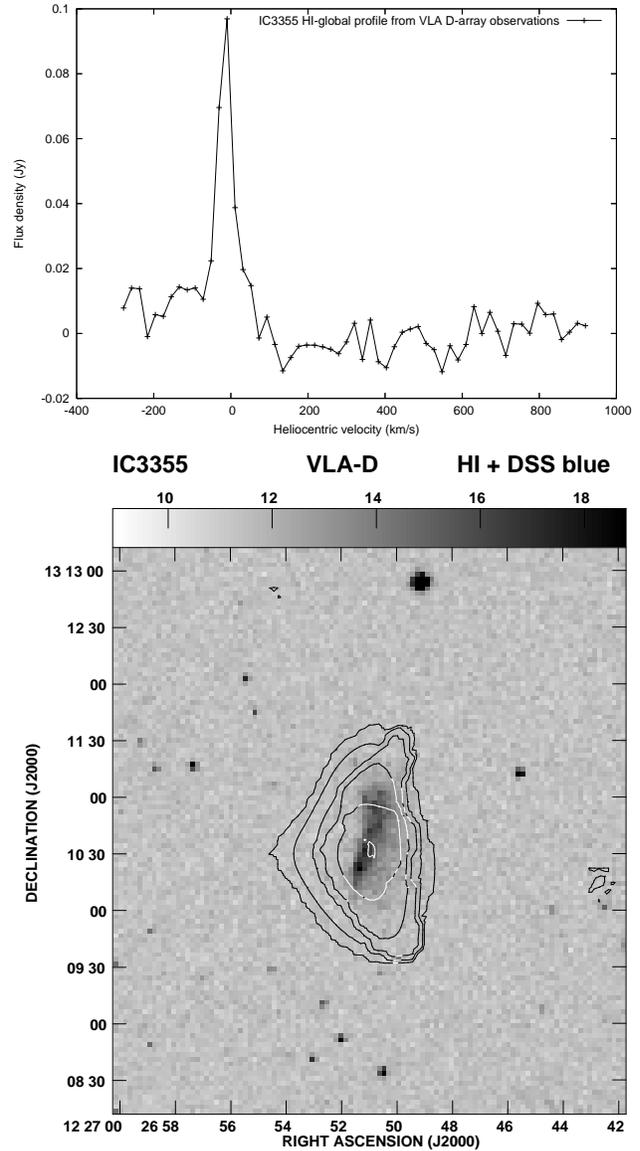,width=3.3in,angle=-90}
  \psfig{file=3355VDBMM0DBL.PS,width=3.3in,angle=0}
   }
\caption[]{Upper panel: The primary beam corrected H{\sc i} emission spectrum of IC~3355 made from a tapered
           VLA D-array image with an angular resolution of $\sim$125 arcsec. Lower panel:  
          The total-intensity H{\sc i} emission contour of IC~3355, observed with the VLA D-array
           with an angular resolution of $\sim$53 arcsec superimposed on the DSS blue-band image.
           Contour levels: (3.84, 5.45, 7.69 $\ldots$)$\times$10$^{19}$ atoms cm$^{-2}$ in steps of $\sqrt{2}$.
}
\end{figure}

The channel maps (Fig. 10) from the VLA D-array data clearly show the detection of
H{\sc i} emission at the position of this galaxy, as noted in Sect.~4.2.  The HI emission
is centered at a position of
RA 12$^h$ 26$^m$ 51.$^s$1, Dec +13$^\circ$ 10$^{\prime}$ 33$^{\prime\prime}$,
which is coincident with the position of the optical galaxy:
RA 12$^h$ 26$^m$ 51.$^s$13, Dec  +13$^\circ$ 10$^{\prime}$ 32.$^{\prime\prime}$6
(Yasuda, Ohamura \& Fukugita 1995). In Fig. 10,  H{\sc i} emission is detected in the
channels with velocities of 11.1, $-$9.5 and $-$30.1 km s$^{-1}$, but no emission is
detected in the channels with velocities of 176.1 and 155.5 km s$^{-1}$, which are
closest to the velocity system at 162 km s$^{-1}$.

At this resolution of $\sim$125 arcsec the primary beam corrected total H{\sc i} flux
density is 5.47 Jy km s$^{-1}$ estimated from the global profile (Fig. 12, upper panel) and
corresponds to a total mass of 3.72$\times$10$^{8}$ M$_\odot$ for IC~3355.
Note that this is almost as high as the H{\sc i} mass of NGC~4438, itself
(4.4$\times$10$^{8}$ M$_\odot$, Sect.~4.2), further emphasizing
that the latter galaxy has been heavily stripped of its gas.
A full-resolution VLA D-array image of IC~3355 (Fig 12, lower panel) shows the emission to be elongated
approximately in the north-south direction, similar to that of the galaxy and consistent
with the structure reported by Cayatte et al.(1990) with an angular resolution of
21$\times$17 arcsec$^2$ along a PA of 75$^{\circ}$ obtained by combining VLA C$-$ and D$-$array data.
The western side of the H{\sc i} gas a sharp gradient but with the emission on the northern
and southern extremeties bending marginally towards the west. The image presented
by Li \& van Gorkom (2001) detects extended emission on the western side which is not
seen in either our image or that of Cayatte et al., possibly due to lower sensitivity.
The detailed structure of the H{\sc i} gas is possibly due to a combination of ram pressure
and tidal interactions since IC~3355 along with NGC~4438, are both likely to be part of the 
M86 sub-cluster, which appears to be merging with the M87 part of the cluster 
(see Kotanyi \& Ekers 1983; Elmegreen et al. 2000; Schindler et al. 1999). 

\subsubsection {An H{\sc i}-tail}
The channel maps at $-$9.5 and $-$30.1 km s$^{-1}$ show extended
diffuse emission which is
seen most prominently at $-$9.5 km s$^{-1}$ in Fig. 10. The global profile of the H{\sc i} tail is
shown in Fig. 13 while the moment 0 image showing the H{\sc i} emission extending
for $\sim$10 arcmin is presented in Fig. 14.  It is of interest to note that the
H{\sc i} emission from the galaxy IC~3355
is also strongest in the velocity channel of $-$9.5 km s$^{-1}$ which is close to its optical
velocity of $-10$ km s$^{-1}$, suggesting that the tail of H{\sc i} emission may be 
of extragalactic origin, although the possibility of it being Galactic foreground emission
cannot be ruled out. If extragalactic, the tail is unlikely to be associated with  
the other companion galaxy, NGC~4435, which has a radial velocity of 801 km s$^{-1}$.
In fact, the latter galaxy, although close in projection to NGC~4438, is likely physically
farther away from NGC~4438, given the large ($\Delta$V =730 km s$^{-1}$) velocity
difference between these two galaxies.

Deep optical images of the Virgo cluster (Phillips \& Malin 1982; Katsiyannis et al. 1998)
show a faint stellar tail extending towards the north of NGC~4438.
Deeper observations trace the
optical tail to a surface brightness of $\mu_{\rm v}\sim$28 mag arcsec$^{-2}$ where it bends
abruptly by $\sim$90$^\circ$ to the west (Mihos et al. 2005).  The `knee' structure of the optical
tail is expected in close and slow encounters in a cluster (cf. Mihos et al. 2005) and is unlikely
to be caused by the interaction between NGC~4438 and 4435, and could be significantly older
than $\sim$100 Myr.  The superposition of our H{\sc i}
image on the deep optical image of Mihos et al. is shown in the bottom panel of Fig. 14.
The H{\sc i} tail we have imaged is close in position and orientation to a
significant part of the optical tail pointing northwards with the H{\sc i} extending till
the bend in the optical tail. The northern tip of the tail is close to the edge of the
half-power point of the primary beam; it would be useful to image the entire field 
to determine whether the H{\sc i} tail follows the optical one beyond the bend.  This could
also provide insights towards understanding whether the tail might be of extragalactic origin.  

Extragalactic H{\sc i} tails could be debris of tidal interaction, ram pressure stripped gas or 
intergalactic H{\sc i} in the cluster medium (see Bekki, Koribalski \& Kilborn 2005).
The tail of H{\sc i} emission in NGC~4388 extends
for $\sim$110$-$125 kpc and has a dynamical time scale of $\sim$100 Myr (Oosterloo \& van Gorkom 2005).
The line flux of the NGC4438 H{\sc i} tail in the velocity channel of $-$9.5 km s$^{-1}$ is
2 Jy km s$^{-1}$ yielding a mass of $\sim$1.36$\times$10$^{8}$ M$_\odot$ if it is at the distance
of the Virgo cluser, which is about 30 per cent
of the H{\sc i} mass of NGC~4438 or 36 per cent of the H{\sc i} mass of IC~3355. For comparison
the mass of the H{\sc i} tail in NGC4388 is 3.4$\times$10$^{8}$ M$_\odot$ (Oosterloo \& van Gorkom 2005).

\begin{figure}
\vbox{
  \psfig{file=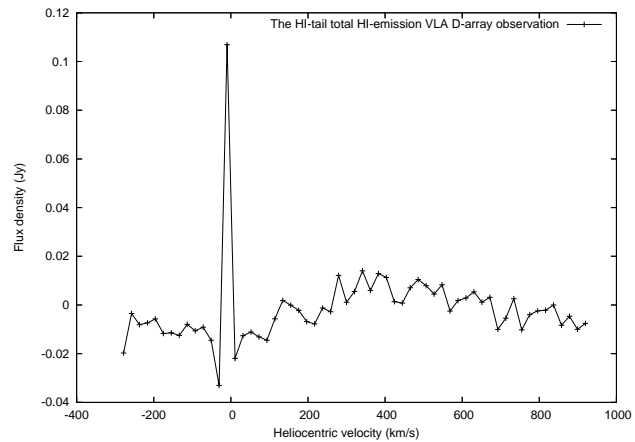,width=3.3in,angle=-90}
   }
\caption[]{
          The primary beam corrected H{\sc i} emission spectrum of the H{\sc i} tail made from a tapered
           VLA D-array image with an angular resolution of $\sim$125 arcsec.
           }
\end{figure}

\begin{figure}
\vbox{
  \psfig{file=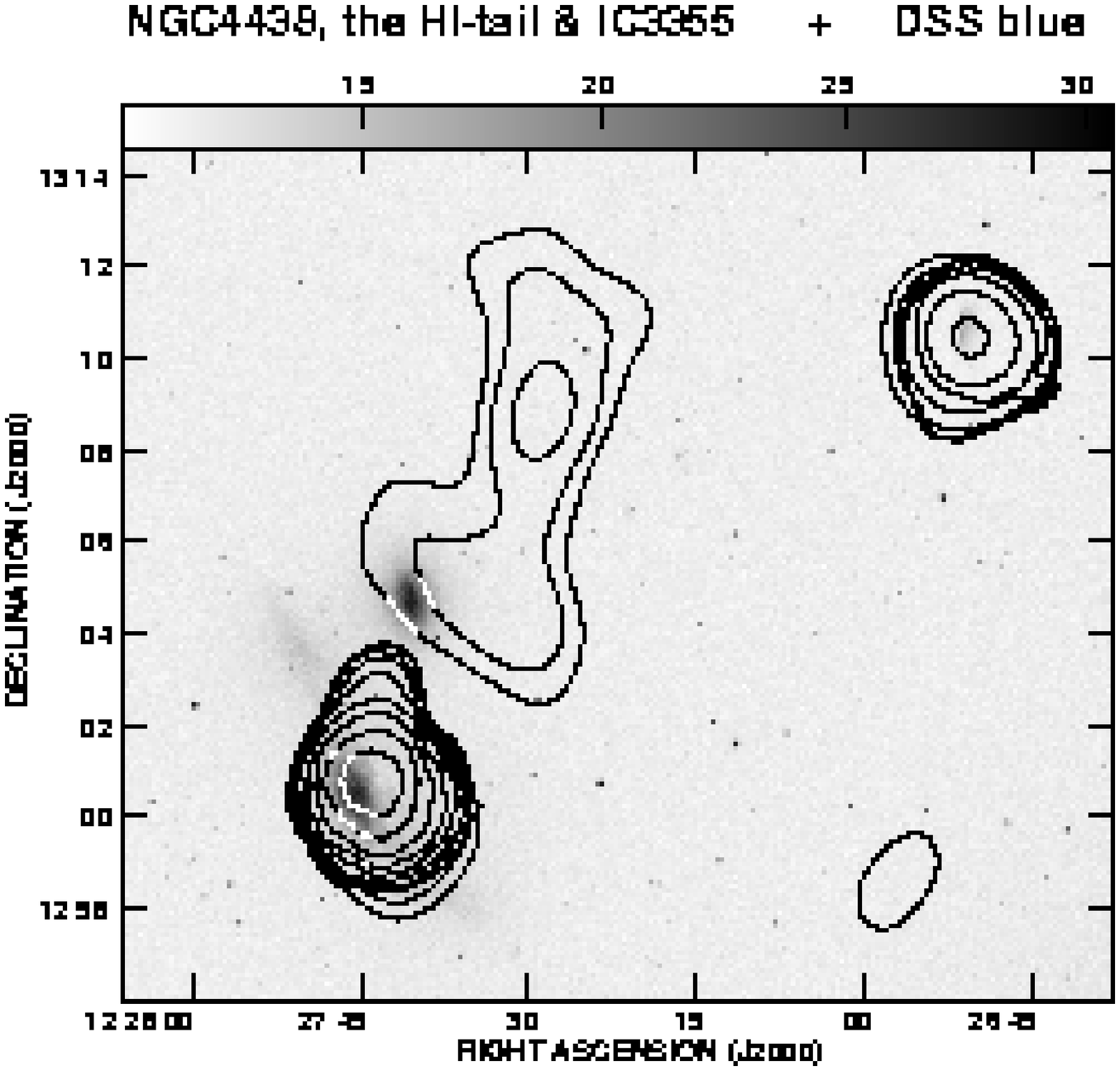,width=3.3in,angle=0}
  \psfig{file=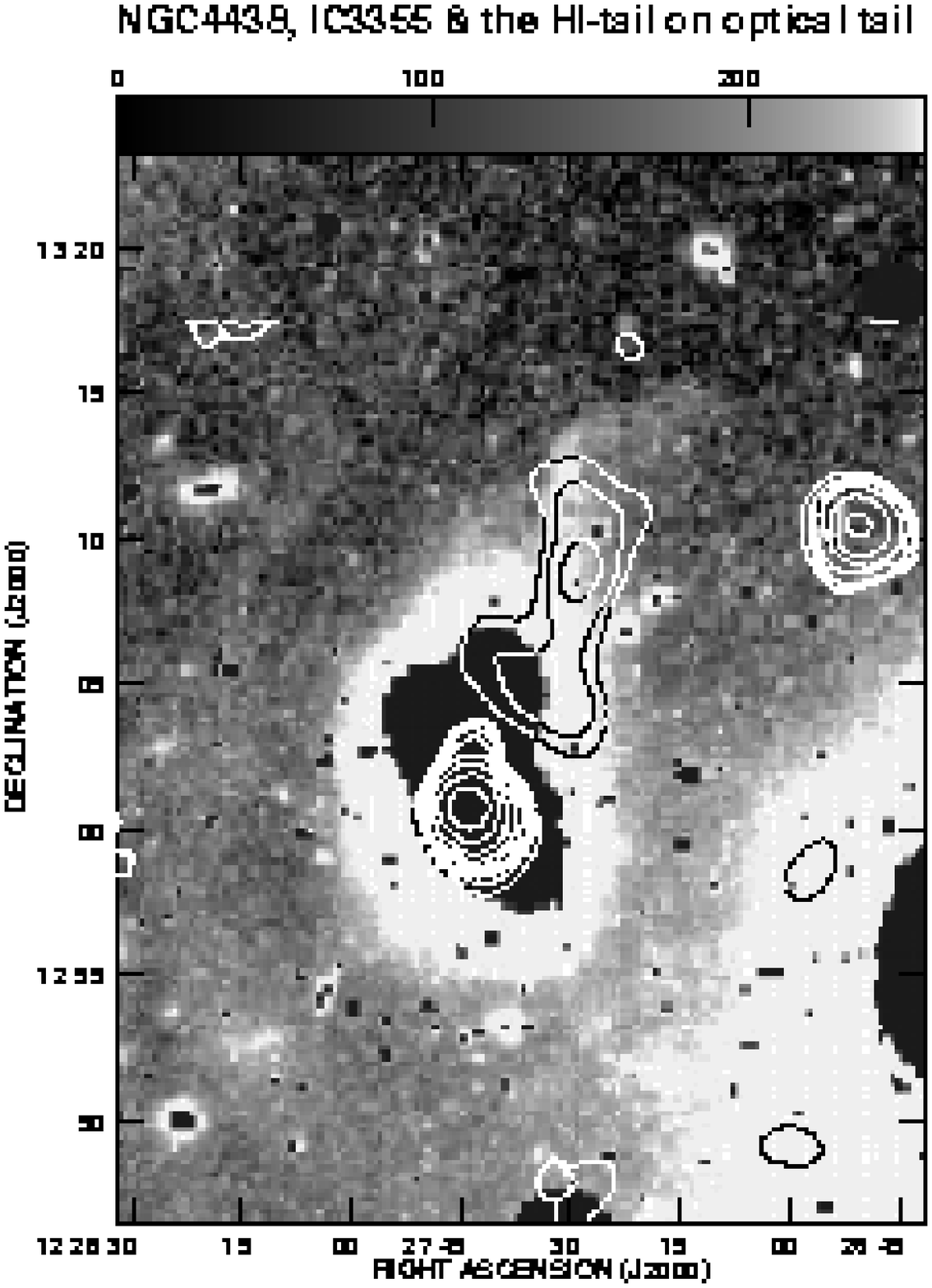,width=3.3in,angle=0}
   }
\caption[]{ The total-intensity contours of H{\sc i} emission of the
           possible tail and the galaxies NGC~4438 and IC~3355 superimposed on
           the DSS blue-band image (upper panel) and the deep optical image 
           (lower panel) from Mihos et al. (2005). The H{\sc i}
           contours levels are (1.13, 1.60, 2.27 $\ldots$)$\times$10$^{19}$ atoms cm$^{-2}$ 
           in steps of $\sqrt{2}$. 
}
\end{figure}

\section{Discussion}

\subsection{The radio-continuum emission}
Our detection of an inverted-spectrum nuclear component suggests that the
small double-lobed radio source with a scale size of $\sim$1 kpc is due to an
AGN rather than a starburst, consistent with earlier suggestions from x-ray
observations (Machacek et al. 2004). 
Although optically thick thermal emission from compact sources with sizes of
$\sim$5$-$10 pc and electron densities of a few thousand cm$^{-3}$ can have an 
inverted spectrum at cm wavelenghts, as seen in the compact star clusters in He 2-10 
(Kobulnicky \& Johnson 1999), the star formation rate in the nucleus of NGC4438 is
modest ($\sim$0.05$-$0.1 M$_\odot$ yr$^{-1}$) and estimates of the electron density
in the nuclear region are small, $\lapp$10 cm$^{-3}$ (e.g. Kenney et al. 1995; 
Machacek et al. 2004).  From a compilation of the structures of nearby
galaxies with a starburst and/or an AGN, Hota \& Saikia (2006) have suggested that such
bubble- or lobe-like radio structures are more likely to be seen in sources with an AGN
rather than a starburst.  

One of the most striking aspects of the structure on scales of $\sim$1 kpc is the 
radio lobes and that the western lobe is clearly seen as a shell-like structure.
The shell-like structures are also seen in H$\alpha$ and
x-ray wavelengths on opposite sides of the nucleus and are closely related to the radio
structure. The H$\alpha$ and x-ray emission possibly arise in regions where the
bubbles of synchrotron-emitting plasma interact with the surrounding ISM.

The radio lobes are very asymmetrically located with the ratio of separations
of the eastern lobe from the nucleus to that of the western one being $\sim$3, while
the corresponding flux density ratio is $\sim$0.60 and 0.57 at 1.4 and 5 GHz respectively.
A higher density on the western side where the ISM of NGC~4438 appears displaced could
provide a viable explanation of the asymmetry in location relative to the nucleus
(cf. Kenney \& Yale 2002).
The lobe on the denser side is also expected to be more luminous due to a higher
efficiency of conversion of beam energy into radio emission, and better confinement
of the radio lobe (cf. Eilek \& Shore 1989; Jeyakumar et al. 2005). However, this simple 
scenario does not explain why the western lobe, which is closer to the nucleus and 
brighter, is also larger in size by a factor of $\sim$2.

A possible solution to this inconsistency comes from the results of numerical
simulations of propagation of jets.  Hydrodynamic simulations of light,
large-scale jets in a decreasing density profile show that the jet bow shock undergoes
two phases, first a nearly spherical one and secondly the well-known
cigar-shaped one (Krause 2002; Krause \& Camenzind 2002; Carvalho \& O'Dea 2002). 
The shell-like structure of the western lobe is suggestive of the first phase of the
development of the bow-shock. In this scenario, the western jet has not
yet entered the cigar phase, and deposits its radio-emitting plasma
in a bigger part of the bubble, almost filling the region within the
bounds of the bow shock. On the other hand, the eastern jet appears
to be in the cigar phase and should therefore have a fairly regular
backflow around it, which flows back into the central parts diffusing and
mixing with the shocked external gas. A similar explanation has been suggested for
the more distant radio galaxy with a starburst, 3C459, which exhibits a similar
asymmetry in the location and structure of the lobes (Thomasson, Saikia \& Muxlow 2003).

For a galaxy moving with a velocity of $\sim$1000 km s$^{-1}$ through the
Virgo cluster, ram pressure could also affect the observed structure of the
source. Using plausible values from Combes et al. (1988) and Vollmer et al. (2005)
the ram pressure is approximately few times 10$^{-11}$ dynes cm$^{-2}$.
For comparison the pressure in the eastern and western lobes are $\gapp$10$^{-10}$
dynes cm$^{-2}$, which corresponds to a cosmic ray energy density of 
$\sim$200 eV cm$^{-3}$ (see Fig. 3a). This is  
significantly larger than the ram pressure, while the
pressure in the interlobe region is comparable to that of ram pressure. The lobes
of emission are unlikely to be significantly affected by ram pressure due to
the motion of the galaxy.

The extended emission displaced to the western side of the disk of the galaxy has a 
spectral index of $\sim$$-$1.1 between 600 and 1400 MHz significantly
steeper than the high-frequency spectrum which is likely to be between 0 and $\sim$$-$0.4. 
This suggests that the 
low-frequency spectral index is dominated by non-thermal emission, while at high-frequencies 
contributions from thermal emission become important. This extended emission is 
unlikely to be due to the AGN. It is also interesting to note that radio emission
has not been detected from the disk of the galaxy; the extended emission possibly
represents plasma which has been displaced from the disk along with
other components of the ISM. 

\subsection{The H{\sc i} gas}
The GMRT and VLA observations show the structure of H{\sc i} emission
from both the disk and extra-planar gas associated with NGC~4438.  
H{\sc i} emission has also been detected 
from IC~3355 as well as a `tail' of H{\sc i} gas extending for $\sim$10$^\prime$ between
NGC~4438 and IC~3355.  

The extra-planar H{\sc i} gas in NGC~4438 has an approximately linear structure with
a total size of $\sim$9.8 kpc and displaced from the disk by  $\sim$4.1 kpc.
The elongated H{\sc i} structure overlaps with the CO extra-planar emission but has a
larger length than the CO feature which has been reproduced in simulations involving both
tidal interactions and ram pressure stripping by the ICM of the Virgo cluster with ISM-ISM
interactions between NGC~4438 and NGC~4435 playing a relatively minor role (Vollmer et al. 2005).
There is a hint that the isovelocity contours in the extra-planar gas appear to initally diverge
while at larger heights they curve inwards. These features, which require confirmation from 
higher spatial resolution observations with adequate surface-brightness sensitivity, are 
reminescent of the galaxy NGC~4522 also in the Virgo cluster, which has been interpreted to be due 
to ram-pressure stripping.  The VLA D-array full-resolution image shows a region
of closed velocity contours where the gas appears to deviate from the overall rotation and
approaches us. A similar feature is also seen in the H$\alpha$ observations at a similar
location but with a much higher approaching velocity (Chemin et al. 2005). There is an x-ray
source at this location and it is tempting to speculate that the outflowing gas may be related 
to energy released from the x-ray source. The high velocity dispersion of the H{\sc i} gas in 
this location would be consistent with such an interpretation.

The faint H{\sc i}-tail we detect towards the north of NGC~4438 is close to a stellar
tail seen earlier in deep optical observations. The  velocity of the H{\sc i} tail is 
$\sim$$-$10 km s$^{-1}$ similar to that of H{\sc i} emission from IC~3355. Although these
aspects suggest that the tail may be in the Virgo cluster, the possibility that it may be a
Galactic feature cannot be ruled out. If it is at a distance similar to that of NGC~4438,
the H{\sc i}-tail has a total extent of $\sim$50 kpc and a mass of 1.4$\times$10$^8$ M$_\odot$. 
It is relevant to note that in NGC~4388 which has an H{\sc i} tail  which extends for $\sim$110 kpc
(Oosterloo \& van Gorkom 2005), there is no evidence of optical light along the length of
the tail except for faint streamers close to the disk. 
Although such H{\sc i} tails could be debris of tidal interaction, ram pressure 
stripped gas or intergalactic H{\sc i} in the Virgo cluster medium (see Bekki, Koribalski 
\& Kilborn 2005), the presence of optical emission light would suggest that it is likely 
to be due to tidal interactions. It would be interesting to make deep H{\sc i} observations 
of NGC~4438 to image this tail beyond the `knee' of the optical tail and also 
find counterparts of the other faint optical tails and arc-like structures listed by 
Mihos et al. (2005).
 
\section{Conclusions}

We have made the most extensive radio continuum and
H{\sc i} observations of the Virgo Cluster galaxy, NGC~4438, and its environment
over a variety of spatial resolutions using the GMRT and the VLA.  We have presented
new data as well as archival data which have been imaged using improved 
data-reduction parameters, revealing new features related to this galaxy.
The radio continuum data span spatial resolutions of 0.2 to 33 arcsec and the
H{\sc i} data span spatial resolutions from 35 to 125 arcsec.

From our high-resolution, radio-continuum data, it is clear that the galaxy harbours
an AGN whose position agrees with the optical and IR nucleus and which has an inverted
radio spectrum.  This confirms ealier speculations, based on x-ray data, that an AGN is present
and is consistent with the presence of radio lobes in the galaxy which are more likely
to be associated with an AGN rather than a starburst (Hota \& Saikia 2006).

We have presented the clearest view of the mini-double-lobed radio source in NGC~4438
and presented the first maps of the minimum energy parameters in the radio lobes.
Note that this is one of
the few spiral galaxies known that contains a clear double-lobed radio source.  
The western lobe extends to $\sim$ 230 pc from the nucleus and the eastern lobe to
$\sim$ 730 pc from the nucleus.  The
lobes break up into shells at higher resolution, visible in x-ray and H$\alpha$
observations as well.  In our radio data, the western shell is clearly delineated.

At lower resolution, a large radio continuum extension is visible, from the 617 MHz
GMRT data, on the south-western side of
the galaxy.  This feature extends as far as 10 kpc from the nucleus of the galaxy.
An analysis of the spectral index of this feature suggests that in addition to the largely
non-thermal emission it may contain a contribution from thermal gas.

The H{\sc i} observations show emission from the disk of the galaxy for the first time
as well as the detailed velocity structure of the disk and extraplanar gas. 
The extraplanar gas appears displaced to the west of NGC~4438 by $\sim$4.1 kpc and
has a mass of $1.8\,\times\,10^8$ M$_\odot$ which is $\sim$1.5 times higher than
the mass in the disk. The extraplanar gas appears to be in regular
rotation and may be rotating more slowly than the gas in the disk. The 
extraplanar gas appears to be rotating about a velocity of $\sim$110 km s$^{-1}$, which is
redshifted compared with the systemic velocity of the galaxy. The velocity structure is 
possibly affected by ram pressure and is reminescent of the
classic ram pressure stripped galaxy NGC4522 in the Virgo cluster.

Our observations also reveal a large tail of H{\sc i} extending north of NGC~4438.
Its velocity is very similar to that of the irregular companion galaxy IC~3355. 
Also the H{\sc i} tail is close to that of a faint optical tail seen in deep images,
suggesting that the tail may be in the Virgo cluster although the possibility that it
might be Galactic emission cannot be ruled out. If it is located close to NGC~4438,
its mass is $1.36\,\times\,10^8$ M$_\odot$, it has a total size of $\sim$50 kpc and 
extends for $\sim$75 kpc from NGC~4438. The possible association of starlight suggests
that it may have formed tidally due to interactions with one of the galaxies in the
M86 merging sub-cluster.   

\section*{Acknowledgments} 
We thank our reviewer, Jeff Kenney, for his comments and criticisms which have 
helped to improve both the scientific content and presentation of the paper. 
AH thanks the Kanwal Rekhi Career Development Scholarship for partial financial support.  
We thank the staffs of GMRT and VLA for making these observations and the VLA staff
for maintaining an excellent archive.
The GMRT is a national facility operated by the National Centre for Radio Astrophysics
of the Tata Institute of Fundamental Research. The VLA is a operated 
by Associated Universities, Inc. under contract with the National Science 
Foundation. This research has made use of the NASA/IPAC extragalactic database 
(NED) which is operated by the Jet Propulsion Laboratory, Caltech, under 
contract with the National Aeronautics and Space Administration. 
We acknowledge the usage of the HyperLeda (http://leda.univ-lyon1.fr) and
GOLDmine (http://goldmine.mib.infn.it) data bases.
JAI wishes to acknowledge a grant from the Natural Sciences and Engineering Research
Council of Canada.\\

\end{document}